%% file: 2019cad.tex
\documentclass[fleqn,usenatbib]{mnras}

\usepackage{newtxtext,newtxmath}

\usepackage[T1]{fontenc}

\DeclareRobustCommand{\VAN}[3]{#2}
\let\VANthebibliography\thebibliography
\def\thebibliography{\DeclareRobustCommand{\VAN}[3]{##3}\VANthebibliography}

\usepackage{graphicx}	
\usepackage{amsmath}	
\usepackage{multicol}        
\usepackage{bm}		
\usepackage{pdflscape}	
\usepackage{graphicx}	
\usepackage{pdflscape}	
\usepackage{threeparttable,lscape}
\usepackage{natbib}
\usepackage{rotating}
\usepackage{color}
\usepackage{xcolor}
\usepackage{appendix}
\usepackage{multirow, array}
\usepackage{longtable}

\newcommand{\Msun}{M$_\odot$}
\newcommand{\nodata}{\centering\arraybackslash --} 
\newcommand{\kms}{km s$^{-1}$}
\newcommand{\cofs}{$^{56}$Co}
\newcommand{\nifs}{$^{56}$Ni}
\newcommand{\orcid}[1]{\href{https://orcid.org/#1}{\includegraphics[width=9pt]{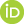}}}

\definecolor{yaleblue}{rgb}{0.1,0.3,0.9}
\definecolor{ultramarine}{rgb}{0, 0, 150}
\definecolor{bostonuniversityred}{rgb}{0.8, 0.0, 0.0}
\definecolor{lava}{rgb}{0.81, 0.06, 0.13}
\definecolor{forestgreen}{rgb}{0.0, 0.27, 0.13}
\hypersetup{colorlinks=true, linkcolor=lava, urlcolor=forestgreen, citecolor=yaleblue}

\title[SN~2019cad: another SN~2005bf-like object]{The double-peaked type Ic Supernova 2019cad: another SN~2005bf-like object}

\author[Guti\'errez et al.]{
\parbox{\textwidth}{
\Large
C.~P.~Guti\'errez$^{\orcid{0000-0003-2375-2064},1,2,3,}$\thanks{E-mail: claudia.gutierrez@utu.fi},
M.~C.~Bersten,$^{4,5,6}$
M.~Orellana,$^{7,8}$
A.~Pastorello,$^{9}$
K.~Ertini,$^{4,5}$
G.~Folatelli,$^{4,5,6}$
G.~Pignata$^{\orcid{0000-0003-0006-0188},10,11}$, 
J.~P.~Anderson,$^{\orcid{https://orcid.org/0000-0003-0227-3451}, 12,11}$
S.~Smartt,$^{13}$ 
M.~Sullivan,$^{\orcid{0000-0001-9053-4820},3}$ 
M.~Pursiainen,$^{3,14}$
C.~Inserra$^{\orcid{0000-0002-3968-4409},15}$,
N.~Elias-Rosa,$^{\orcid{0000-0002-1381-9125},9,16}$
M.~Fraser$^{\orcid{0000-0003-2191-1674},17}$, 
E.~Kankare$^{\orcid{0000-0001-8257-3512},2}$,
M.~Stritzinger$^{\orcid{0000-0002-5571-1833},18}$,
J.~Burke,$^{19,20}$
C.~Frohmaier,$^{21}$
L.~Galbany$^{\orcid{0000-0002-1296-6887},22}$,
D.~Hiramatsu$^{\orcid{0000-0002-1125-9187},19,20}$,
D.~A.~Howell$^{\orcid{0000-0003-4253-656X},19,20}$,
H.~Kuncarayakti,$^{2,1}$
S.~Mattila,$^{2}$
T.~M\"uller-Bravo$^{\orcid{0000-0003-3939-7167},3}$,
C.~Pellegrino,$^{19,20}$
M.~Smith$^{23}$
\\
\textit{Affiliations are listed at end of paper}
}
}

\date{Accepted XXX. Received YYY; in original form ZZZ}

\pubyear{2021}

\begin{document}
\label{firstpage}
\pagerange{\pageref{firstpage}--\pageref{lastpage}}
\maketitle

\begin{abstract}
We present the photometric and spectroscopic evolution of supernova (SN) 2019cad during the first $\sim100$ days from explosion. Based on the light curve morphology, we find that SN~2019cad resembles the double-peaked type Ib/c SN~2005bf and the type Ic PTF11mnb. Unlike those two objects, SN~2019cad also shows the initial peak in the redder bands. Inspection of the $g-$band light curve indicates the initial peak is reached in $\sim$8 days, while the $r$ band peak occurred $\sim$15 days post-explosion. A second and more prominent peak is reached in all bands at $\sim$45 days past explosion, followed by a fast decline from $\sim$60 days. During the first 30 days, the spectra of SN~2019cad show the typical features of a type Ic SN, however, after 40 days, a blue continuum with prominent lines of \ion{Si}{ii} $\lambda6355$ and \ion{C}{ii} $\lambda6580$ is observed again. Comparing the bolometric light curve to hydrodynamical models, we find that SN~2019cad is consistent with a pre-SN mass of 11 \Msun, and an explosion energy of 3.5$\times 10^{51}$ erg. The light curve morphology can be reproduced either by a double-peaked \nifs\ distribution with an external component of 0.041 \Msun, and an internal component of 0.3 \Msun \, or a double-peaked \nifs\ distribution plus magnetar model ($P\sim$11~ms and $B\sim$26$\times 10^{14}$~G). If SN~2019cad were to suffer from significant host reddening (which cannot be ruled out), the \nifs\ model would require extreme values, while the magnetar model would still be feasible.
\end{abstract}




\section{Introduction}

Core-Collapse Supernovae (SNe) are produced by the explosion of massive stars (M$_{ZAMS}>8-10$ \Msun). They are traditionally classified into different classes depending on the presence or absence of certain lines. SNe from collapsing stars that do not show hydrogen but show helium in their spectra are classified as type Ib SNe (SNe~Ib), while those with not hydrogen or helium are classified as Ic SNe (SNe~Ic; e.g., \citealt{Filippenko97, GalYam17, Modjaz19}). The absence of these spectral lines implies their progenitor stars shed their hydrogen- and helium-rich envelops over their lifetimes, mainly due to strong stellar winds \citep{Heger03,Georgy09} or binary interaction \citep[e.g.,][]{Podsiadlowski92,Nomoto95,Eldridge08}. While both mechanisms are successful in explaining the absence of the hydrogen layer, removing the helium layer is still a challenge as it is found in denser parts of the star. Given these complications, it has been proposed that some helium is possibly present in SNe~Ic but it is not seen in the spectrum because it is not excited (\citealt{Dessart12a}, but see \citealt{Williamson20}). On the other hand, recent observational evidence, such as the low progenitor masses inferred for SNe~Ic from their light curves \citep{Drout11,Lyman16,Taddia18} and the relative SN~Ic rate \citep{Smith11}, favour the the binary scenario.  

The direct identification of stars in pre-explosion images can give more insights about the nature of the SN progenitor \citep[e.g.][]{VanDyk02,VanDyk03,Mattila08,Smartt09,VanDyk17}. However, for hydrogen-free objects only a couple of cases exist. A confirmed progenitor for the SN~Ib iPFT13bvn \citep{Cao13,Folatelli16} and two progenitor candidates, one for the type Ic SN~2017ein \citep{VanDyk18,Kilpatrick18,Xiang19} and another one for the type Ib SN~2019yvr \citep{Kilpatrick21}. If post explosion images confirm the progenitor association, then it would represent the first progenitor detection for a SN~Ic and the second for a SN~Ib.

SNe~Ic are the most intriguing objects among core-collapse events. They represent a heterogeneous class showing a large range in luminosity and light-curve shapes \citep[e.g.,][]{Bianco14,Lyman16,Prentice16,Taddia18}, as well as diverse spectra \citep[e.g.,][]{Matheson01,Modjaz14,Shivvers19}. A small fraction of SNe~Ic have been observed with broad absorption lines that have been associated with long-duration gamma-ray bursts (GRBs; \citealt{Woosley06,Cano17}). These objects, usually labelled as broad line SNe Ic (SNe~Ic-BL), represent a challenge in our understanding of the explosion mechanism and final steps of massive-star evolution. Fortunately, high-cadence surveys are detecting and following up more objects of this type allowing their characterisation and physical understanding.  

Stripped-envelope SNe, considered as hydrogen-deficient objects, usually present bell-shaped light curves, with a single peak reached a couple of weeks after explosion. These light curves are powered by the decay of \nifs\ to \cofs, and then to $^{56}$Fe. Some stripped-envelope SNe with a well-constrained explosion date and good photometric cadence have shown early emission prior to the usual nickel peak. When it is observed, this initial peak lasts a few days and has been attributed to the cooling of the ejecta after the shock breakout \citep[e.g.,][]{Woosley94, Bersten12,Nakar14}, while the second peak has a duration of a couple of weeks and is mainly powered by the decay of \nifs. There are now many cases where the SN was observed before the nickel peak \citep[see][and references therein]{Modjaz19}, but most of them were classified as SN~IIb (transitional objects between SNe~II and SNe~Ib) and the emission can be well explained as due to the cooling of a thin but extended hydrogen-envelope. In the other known cases, they showed some peculiarities, as was the case of the type Ib SN~2008D associated with an X-ray flash \citep{Soderberg08,Mazzali08}, or the cases of the SNe~Ic SN~2006aj \citep{Campana06} and more recently SN~2017iuk \citep{Izzo19}  associated with long-duration GRBs (i.e. SN~Ic-BL). In these cases, the early emission is harder to explain as due to the cooling of an envelope because of the compact nature of their progenitor. Some alternatives, such as the presence of circumstellar material, the cooling of a cocoon jet or some external nickel seem to be required \citep{Soderberg08,Bersten13}.

Unlike the early emission discussed above, which lasts for a few days, there are two objects in the literature where the early emission is longer in duration, appearing as a peak at around 20 days from the explosion, followed by a main peak occurring at $\sim40$ days from the explosion. These objects are SN~2005bf \citep{Anupama05, Tominaga05, Folatelli06} and PTF11mnb \citep{Taddia18a}. SN~2005bf was classified as a transitional object between SNe~Ic and SNe~Ib \citep{Folatelli06}, while PTF11mnb was cataloged as a typical SN~Ic. \citet{Anupama05} claimed that SN~2005bf was the explosion of a massive He star with some H left. \citet{Tominaga05} concluded that the progenitor was a Wolf-Rayet WN star and the morphology of the light curve can be reproduced by a double-peaked \nifs\ distribution. \citet{Folatelli06} found that the most favored model is consistent with an energetic and asymmetric explosion of a massive WN star, where an unobserved relativistic jet was launched producing a two-component explosion. Finally, through analysing late phase spectroscopy and photometry of SN~2005bf, \citet{Maeda07} suggested that the power source of this SN is a magnetar. On the other hand, for PTF11mnb, \citet{Taddia18a} suggested that the progenitor was a massive single star with a double-peaked \nifs\ distribution powering the SN light curve.

In this paper, we present photometry and spectra of a peculiar event, the type Ic SN~2019cad. We discuss its observed properties and compare it with SN~2005bf and PTF11mnb. The paper is organized as follows. A description of the observations and data reduction are presented in Section~\ref{sec:obs}.
In Section~\ref{sec:ana}, we describe the photometric and spectral properties of SN~2019cad and a comparison with similar events. The progenitor properties are analysed through hydrodynamic modeling in Section~\ref{sec:model}, while in Section~\ref{sec:disc} we present the discussion and conclusions. 
Throughout, we assume a flat $\Lambda$CDM universe, with a Hubble constant of $H_0=70$\,km\,s$^{-1}$\,Mpc$^{-1}$, and $\Omega_\mathrm{m}=0.3$.

\section{Observations of SN~2019cad}
\label{sec:obs}

SN~2019cad (also known as ZTF19aamsetj and ATLAS19ecc) was discovered by the Zwicky Transient Facility \citep[ZTF;][]{Bellm19, Graham19} on 2019 March 17 (MJD$=58559.24$) at a magnitude of m$_r=19.02\pm0.11$ mag \citep{Nordin19cad}. The last non-detection obtained by ZTF occurred on 2019 March 16 (MJD$=58558.19$) with a detection limit of m$_r\sim19.183$ mag. A deeper early detection and a non-detection were obtained by the Asteroid Terrestrial-impact Last Alert System (ATLAS; \citealt{Tonry18, Smith20}) on 2019 March 12 (MJD$=58554.42$; m$_o=19.91\pm0.11$ mag) and 2019 March 4 (MJD$=58546.46$; m$_o=20.50$ mag), respectively. These new constraints allow us to adopt the midpoint between the last non-detection and first detection as the explosion epoch (MJD$=58550.44\pm4$ (2019 March 8). 

On 2019 March 22 (MJD$=58564.36$), SN~2019cad was observed spectroscopically by the Global SN Project (GSP) and classified as a SN~Ic around maximum light at a redshift $z=0.0267$ \citep{Burke19}. In the classification report, \citet{Burke19} noted the object was slightly faint (i.e., M$_r \sim-16.6$ mag) compared what is typically observed in SNe~Ic. Twenty days after the classification, the $r$ band ZTF light curve showed a rebrightening, which was confirmed 5 days later in the $g$ band. Because of this rebrightening, we started a follow-up campaign. The details of which are now described.

\subsection{Photometry}
\label{slc}

Photometric coverage of SN~2019cad was acquired using different facilities and instruments over a period of 14 weeks as follows:
\begin{itemize}
\item \textbf{ATLAS:} Photometry in the orange ($o$) filter (a red filter that covers a wavelength range of 5600 to 8200 \AA) and cyan ($c$) filter (wavelength range 4200 to 6500 \AA) was obtained by the twin 0.5 m ATLAS telescope system \citep{Tonry18}, spanning from 2019 March 11 to 2019 May 30. The data were reduced and calibrated automatically as described in \citet{Tonry18} and \citet{Smith20}. Table~\ref{photoatlas} lists the mean magnitudes.

\item \textbf{Gran Telescopio Canarias:} One epoch of $r$-band photometry was obtained with the 10.4-m Gran Telescopio Canarias (GTC) using OSIRIS. The $r$-image was reduced with \textsc{iraf} following standard procedures. The photometry was performed using the \textsc{python} package \textsc{photutils} \citep{Bradley19} of \textsc{astropy} \citep{Astropy18}.
The $r$ magnitude was calibrated using Pan-STARRS \citep{Chambers16,Magnier20}.

\item \textbf{Las Cumbres Observatory:} 
Multiband photometry was obtained with the 1.0-m telescopes of Las Cumbres Observatory \citep{Brown13} at three different epochs through the Global Supernova Project (GSP). The data reduction and SN photometry measurements were performed following the prescriptions described in \cite{Firth15}. 

\item \textbf{Liverpool Telescope:} Five epochs of $ugriz$ were obtained with the 2-m Liverpool Telescope \citep[LT;][]{Steele04} using the IO:O imager. LT data were reduced using the standard IO:O pipeline. The photometry was performed using \textsc{photutils}. The $ugriz$ magnitudes were calibrated
using Pan-STARRS sequences.

\item \textbf{Nordic Optical Telescope:} Using the 2.56-m Nordic Optical Telescope (NOT) at Roque de los Muchachos Observatory, we obtained six epochs of $ugriz$ photometry from April 23 to June 19, and three epochs of $JHK$ photometry from April 30 to June 13. The optical observations were performed with ALFOSC, and the near-infrared (NIR) observations with NOTCam. 
All NOT observations were obtained through the NOT Unbiased Transient Survey (NUTS) allocated time. Optical and NIR data reduction and SN photometry measurements were performed using the \textsc{python/pyraf} \textsc{SNOoPY} pipeline \citep{Cappellaro14}. The $ugriz$ magnitudes were calibrated using observations of local Sloan and Pan-STARRS sequences, while the $JHK$ magnitudes were calibrated using 2MASS \citep{Skrutskie06}.

\item \textbf{Neil Gehrels \textit{Swift} Observatory:} One epoch of UltraViolet (UV) Optical observations were obtained with the UltraViolet/Optical Telescope (UVOT) on board the Swift spacecraft. Imaging observations were processed with aperture photometry following \citet{Brown09}. No template subtractions were achieved.

\item \textbf{William Herschel Telescope:} One epoch of $gr$ photometry was obtained with ACAM in the William Herschel Telescope (WHT) on 2019 June 20. The images were reduced with \textsc{iraf} following standard procedures, while the photometry was performed using \textsc{photutils}. The $gr$ magnitudes were calibrated using Pan-STARRS.

\end{itemize}

Optical, NIR and UVOT photometry are presented in Table~\ref{photo}, \ref{photonir} and \ref{photoswift}, respectively. 
Additional photometry in the $gr$ bands was obtained from the ZTF public stream through the Lasair\footnote{\url{https://lasair.roe.ac.uk/}} broker \citep{Smith19} and presented in Table~\ref{photoZTF}.

\subsection{Spectroscopy}
\label{sspect}

SN~2019cad was observed spectroscopically at 12 epochs spanning phases between 13.6 to 88.1 days past explosion. These observations were acquired with five different instruments: ALFOSC at the NOT, SPRAT \citep{Piascik14} at the LT, the FLOYDS spectrograph \citep{Brown13} on the Faulkes Telescope South (FTS), OSIRIS at the GTC, and with the Spectral Energy Distribution Machine \citep[SEDM;][]{Blagorodnova18}
on the automated 60-inch telescope at Palomar Observatory \citep[P60;][]{Cenko06}\footnote{Public spectrum obtained from the TNS webpage: \url{https://wis-tns.weizmann.ac.il/object/2019cad}}. 
The log of spectroscopic observations of SN~2019cad is presented in Table~\ref{tspectra}.

Data reductions for ALFOSC and OSIRIS were performed following standard routines in \textsc{iraf}. The SPRAT data were reduced using the Fast and Dark Side of Transient experiment Fast extraction pipeline (FDSTfast)\footnote{\url{https://github.com/cinserra/FDST}}. FLOYDS spectra were reduced using the \textsc{pyraf}-based \textsc{floyds\_pipeline}\footnote{\url{https://github.com/LCOGT/floyds_pipeline}} \citep{Valenti14}, while the public SEDM spectrum was automatically reduced by the IFU data reduction pipeline \citep{Rigault19}. All spectra are available via the WISeREP\footnote{\url{https://wiserep.weizmann.ac.il/}} repository \citep{Yaron12}.

\section{Characterising SN~2019cad}
\label{sec:ana}

\subsection{Host galaxy}
\label{sec:gal}

\begin{figure}
\centering
\includegraphics[width=\columnwidth]{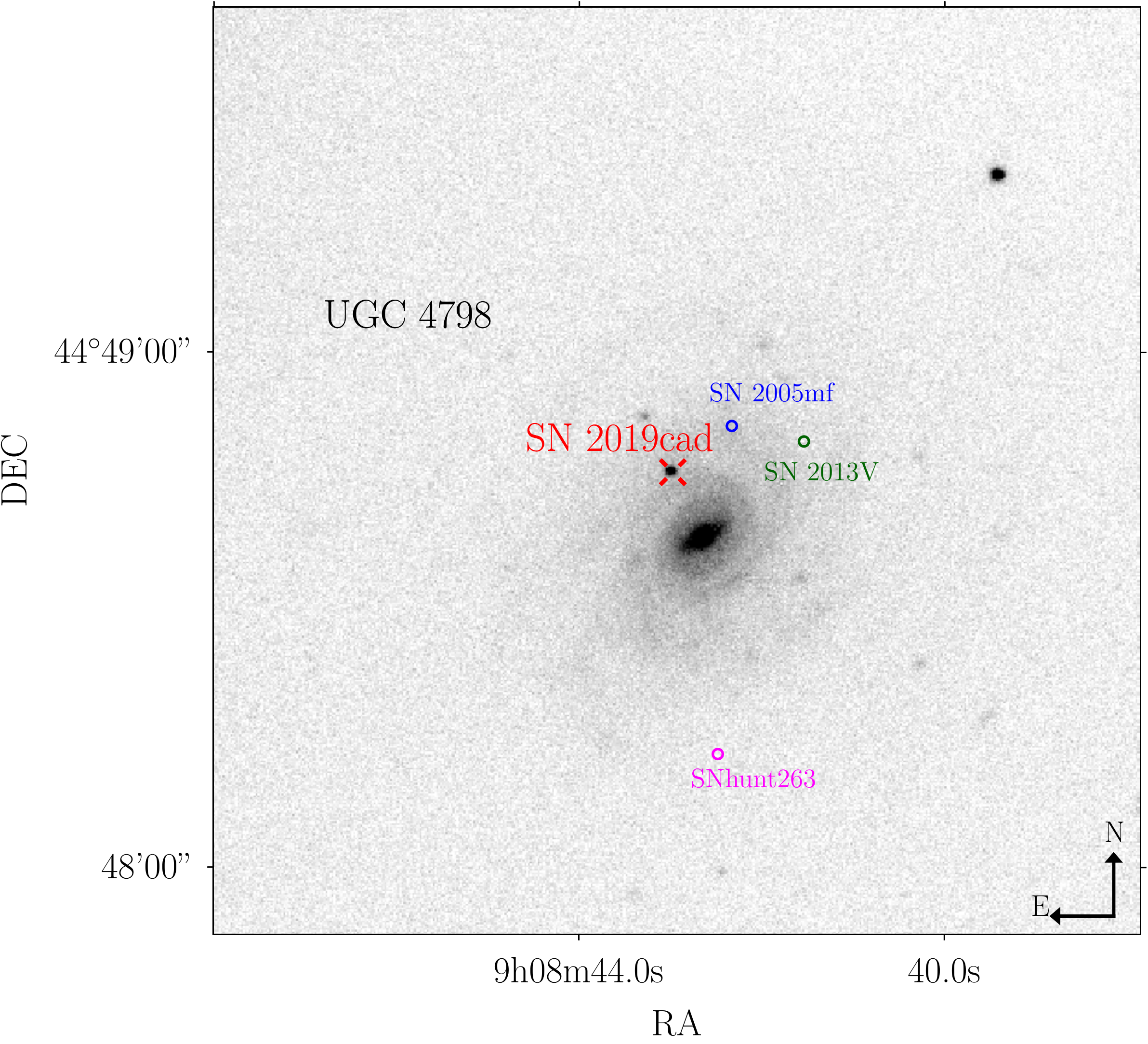}
\caption{NOT $r$-band image of SN~2019cad and its host galaxy. SN~2019cad is marked with a red crosshair. The locations of SN~2005mf (Ic), SN~2013V (Ia), and SNHunt263 (Ia) are marked in blue, green and magenta circles, respectively. The orientation of the image is indicated in the bottom-right corner.
}
\label{galaxy}
\end{figure}

The host galaxy of SN~2019cad was identified as  UGC~4798, a spiral galaxy at a redshift of 0.0267 \citep{deVaucouleurs91}\footnote{\url{http://ned.ipac.caltech.edu/}}. Adopting the recessional velocity corrected for Local Group infall into Virgo reported by HyperLEDA\footnote{\url{http://leda.univ-lyon1.fr}} (\citealt{Makarov14}; $v_{vir}=8152\pm3$ \kms), we obtain a distance modulus $\mu=35.38\pm0.13$ mag, which is equivalent to a distance of $119.2\pm8.4$ Mpc. 

SN~2019cad was located 3{\farcs}6 east and 7{\farcs}6 north from the galaxy centre. Its host has been classified as a \ion{H}{i}-rich galaxy \citep{Wang13G} with a stellar mass of $\log_{10}\left(\mathrm{M}_*\right)=10.59$ \Msun\ \citep{Chen12G} and a star-formation rate (SFR) of $1.97\pm0.01$ \Msun yr$^{-1}$ \citep{Cormier16G}. SN~2019cad is the fourth SN reported in UGC~4798. SN~2005mf was detected at $5{\farcs}9$ west and $13{\farcs}3$ north from the galaxy centre \citep{DSN05mf}, SN~2013V was found at $11{\farcs}5$ west and $11{\farcs}1$ north of the centre of the galaxy \citep{DSN13V}, while SNHunt263 was detected at 1{\farcs} west and 25{\farcs} south of the center of the galaxy \citep{Drake09}. SN~2005mf was classified as a SN~Ic \citep{CSN05mf}, while SN~2013V and SNHunt263 as SNe~Ia \citep{CSN13V,CSNHunt263}. All these SNe have been located at projected distances larger than 4.5 kpc from the centre of the galaxy, see Figure~\ref{galaxy}.
The substantial number of SNe detected in UGC~4798 in the last 15 years is not rare. Previous studies \citep[e.g.][]{Thoene09,Anderson13} have shown that several galaxies have been hosted multiple SNe. Most of these galaxies have a high SFR. However, for NGC~2770, the high number of SNe~Ib was found by \citet{Thoene09} to be a coincidence. 

To estimate the oxygen abundance of UGC~4798, we use the public spectrum of the central region of the galaxy obtained from the Sloan Digital Sky Survey \citep[SDSS;][]{SDSS19}. Applying the O3N2 diagnostic method from \citet{Pettini04}, we obtain an oxygen abundance of $12+\log\left(\mathrm{O/H}\right) = 8.72\pm0.08$. Employing the same method, 
\citet{Modjaz11} reported an oxygen abundance at the position of SN~2005mf of $12+\log\left(\mathrm{O/H}\right) = 8.66\pm0.09$, which is around the solar abundance \citep[e.g.][]{Pettini04}. Based on the small differences in these two estimations, we suggest the oxygen abundance near SN~2019cad to be around solar.

\subsection{Extinction estimation}
\label{sec:extinction}

To determine the intrinsic properties of SN~2019cad, an estimation of the total reddening (from both the Milky Way and the host galaxy) along the line of sight to the object is needed. The galactic reddening was found to be $E(B-V)_{MW}=0.015$ mag \citep{Schlafly11}. To calculate the host galaxy extinction, we searched for the \ion{Na}{i}\,D absorption lines in the SN spectra. A narrow \ion{Na}{i}\,D line at the host galaxy rest wavelength is clearly detected in the NOT spectra, with an equivalent width (EW) of $\sim1.32$ \AA. This strong absorption suggests a significant reddening from the host galaxy. Using the relations of $E(B-V)$ and EW(\ion{Na}{i}\,D) found by \citet{Turatto03} from low and heavily reddened SNe~Ia, we obtain two values: $E(B-V)_{\rm {Host}}=0.21$ and $0.63$ mag. Applying the empirical relation from \citet{Poznanski12}, we find an $E(B-V)_{\rm Host}=0.49\pm0.08$ mag. Unfortunately, there are large discrepancies in these estimations, which likely caused by the fact that the lines have saturated \citep{Munari97,Poznanski11}. Therefore, we explored alternative methods to determine the host extinction.

Constraints on the host extinction can be also found using the colour of the SN around the maximum \citep{Drout11, Taddia15, Stritzinger18b}. \citet{Drout11} found that in SNe~Ibc the $V-R$ colour at 10 days from the $V$-band maximum shows very small scatter, varying between 0.18 and 0.34 mag. Later, \citet{Stritzinger18b}, following a similar approach, but using a range of optical or optical/NIR colour combinations, built intrinsic colour-curve templates, which we can use to measure the colour excess. By comparing the SN~2019cad $g-r$ and $r-i$ colours with the templates from \citet{Stritzinger18b}, we found an excess in $r-i$ that corresponds to $E(B-V)\sim0.22$ mag, however, the $g-r$ colours were found to be bluer than the template (see Figure~\ref{lc}, middle panel). Because of the inconsistency in the colours (excess in $r-i$, but deficit in $g-r$), this method is not applicable in this SN, which could have been expected from its peculiar evolution.

Another method to estimate the host reddening  is to derive the Balmer decrement from an \ion{H}{ii} region that we assume is sharing the same extinction of the SN along the line of sight. In the latest spectrum of SN~2019cad, a weak emission from  H$\alpha$ and H$\beta$ is detectable. To measure the flux of latter emission lines, we remove the SN flux following the procedure presented in \citet{Pignata11}.  We stress the fact that due to the very low signal-to-noise ratio of H$\beta$ ($\sim2.0$), the measurement of its flux is very uncertain, therefore we consider its minimum and maximum flux to obtain an estimation of the maximum ($E(B-V)=0.80$ mag) and minimum ($E(B-V)=0.28$ mag) color excess along the line of sight of SN~2019cad. Even though with large uncertainties, such values provide additional evidence that SN~2019cad is highly reddened.

The reddening estimation is one of the largest systematic uncertainties in the SN field. As we have shown above, large differences are found using different methods. The methods used were obtained for normal core-collapse SNe, and even in these cases it is not clear which methods are robust, if any. The situation is even worse for peculiar events as SN~20019cad. Therefore, we decided not to consider host galaxy extinction in the rest of this paper, but we explore in Section~\ref{sec:comp} and Section~\ref{sec:disc} the implications of a larger host galaxy reddening ($E(B-V)_{Host}=0.49$ mag, the median of these estimations and the value obtained with the most recent empirical relation of $E(B-V)$ and EW(\ion{Na}{i}\,D)).

\subsection{Light curves}
\label{sec:lc}

\begin{figure*}
\centering
\includegraphics[width=0.75\textwidth]{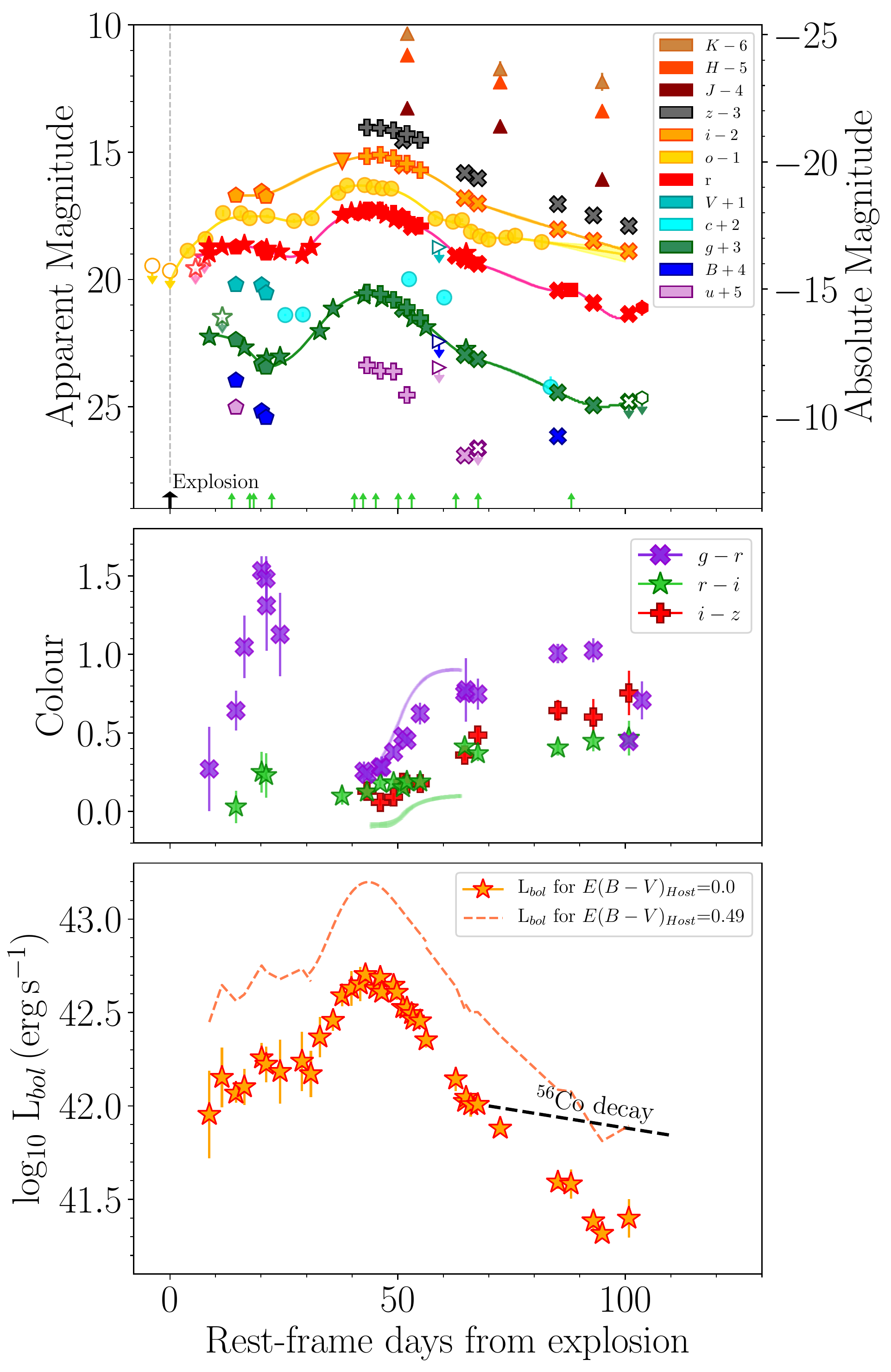}
\caption{\textbf{Upper:} Optical light curves of SN~2019cad. ATLAS photometry is presented as circles, ZTF photometry as stars, NOT photometry as crosses, LT photometry as pluses, Las Cumbres photometry as pentagons, NOTCam photometry as up-facing triangles, GTC photometry as squares, WHT photometry as hexagons, and Swift photometry as right-facing triangles. Upper limits are presented as open symbols. The explosion estimation is indicated as a vertical black arrow. The vertical lime-green arrows represent epochs of optical spectroscopy. The photometry is corrected for Milky Way extinction. The solid lines show the Gaussian process (GP) interpolation. \textbf{Middle:} Colour curves of SN~2019cad. Only corrections for Milky Way extinction have been made. Solid lines (purple and green) show the intrinsic color-curves templates ($g-r$ and $r-i$) from \citet{Stritzinger18b}. \textbf{Bottom:} Bolometric light curve of SN~2019cad assuming an $E(B-V)_{Host}=0.0$ mag (stars) and $E(B-V)_{Host}=0.49$ mag (dashed line).
}
\label{lc}
\end{figure*}

The multiband light curves of SN~2019cad are shown in Figure~\ref{lc} (top panel). SN~2019cad presents an unusual light curve evolution characterised by the presence of an initial peak between 10-20 days from explosion, followed by the main peak at $\sim45$ days. This double-peak light curve resembles those of the peculiar type Ic SN~2005bf \citep{Folatelli06} and PTF11mnb \citep{Taddia18a}.

To estimate the main parameters of the light curves, we use Gaussian processes (GPs). Following \citet{Gutierrez20}, we perform the light curve interpolation with the \textsc{python} package \textsc{george} \citep{Ambikasaran16} using the Matern 3/2 kernel. We find that SN~2019cad reaches an initial peak absolute $g$-band magnitude of ${\rm M}_{g}=-16.35$ mag in $\sim8$ days. In the $r$ and $o$ bands, the initial peak is ${\rm M}_{r}= -16.68$ mag and ${\rm M}_{o}= -16.86$ mag at $\sim15$ and $\sim18.9$ days, respectively. After this peak, the light curves in the $o$ and $r$-bands show a small decrease in brightness (between $\sim0.2-0.4$ mag in $\sim10$ days), while the $g$-band decreases more than 1 mag in the same period ($\sim13$ days). Following this initial peak, a rise of 1.5 -- 3 mag is observed in all bands. The peak in $groi$ occurs at $\sim43.9$, 44.3, 41.5, and 44.9 days with magnitudes of ${\rm M}^{\rm max}_{g}=-17.83$ mag, ${\rm M}^{\rm max}_{r}=-18.10$ mag, ${\rm M}^{\rm max}_{o}=-18.07$ mag, and ${\rm M}^{\rm max}_{i}=-18.26$ mag. Once SN~2019cad has reached the main peak, the light curves decrease by 2.20 ($g$), 1.71 ($r$), 1.67 ($i$), 1.50 ($o$) and 1.45 ($z$) mag in $\sim20$ days. Table~\ref{LCparam} shows a summary of the light-curve parameters.

\begin{table*}
\begin{center}
\caption{Light curve parameters of SN~2019cad assuming an $E(B-V)_{Host}=0$ mag}
\begin{tabular}{cccccccc}
\hline
\hline
Band & Initial Peak & Epoch   & Main Peak  & Epoch & Change in magnitude    & Decline rate        \\
     & magnitude    & (d)     & magnitude  &  (d)  & from the main peak to  & after 60 days       \\
     &  (mag)       &         &  (mag)     &       & 20 d post-peak (mag)   & (mag per 100d) \\
\hline
\hline
$g$  & $-16.35$     & 8       &   $-17.83$ & 43.9  &      2.20              & $7.10\pm0.21$       \\
$r$  & $-16.68$     & 15      &   $-18.10$ & 44.3  &      1.71              & $5.91\pm0.25$       \\ 
$o$  & $-16.86$     & 18.9    &   $-18.07$ & 41.5  &      1.50              & $5.33\pm1.12$       \\
$i$  & \nodata      & \nodata &   $-18.26$ & 44.9  &      1.67              & $5.83\pm0.10$       \\
\hline
\end{tabular}
\label{LCparam}
\end{center}
\end{table*}

From 60 to 100 days, the light curve shows a linear decline in $griz$. The slope of the decline in all bands is faster than the expected from the full trapping of gamma-ray photons and positrons from the decay of \cofs\ \citep[0.98 mag per 100 days;][]{Woosley89}. Fitting a line to the tail, we measure a slope of $7.10\pm0.21$ mag (100d)$^{-1}$ in the $g$-band and $5.91\pm0.25$ mag (100d)$^{-1}$ in the $r$-band. To guarantee that we use luminosities when the light curve has entered the nebular phase, we follow the prescriptions of \citet{Meza20} and we fit a line to the tail from 75 to 100 days. Thus, we derive a slope of $6.54\pm0.31$ mag (100d)$^{-1}$ in $g$ and $6.00\pm0.21$ mag (100d)$^{-1}$ in $r$. \citet{Taddia18} found that the linear decay slope for SNe~Ic are between 1.7 and 2.7 mag (100d)$^{-1}$. These declines are faster than the expected from the \cofs\ decay, but slower than those measured in SN~2019cad.
If the radioactive decay powers the light curve, then the extremely fast decline suggests a significant leakage of gamma-ray photons which can, for example, be achieved assuming a low-mass ejecta.

The colour curves of SN~2019cad are presented in the middle panel of Figure~\ref{lc}. During the first 20 days, SN~2019cad becomes redder, going from a colour of $g-r=0.27$ mag at 8.6 days to $g-r=1.53$ mag at 20.1 days. At around 20 days, the $g-r$ colour shows a maximum that corresponds to the initial peak in the optical light curves. After this peak, the SN evolves to bluer colours, reaching a minimum of $g-r=0.24$ mag at $\sim44$ days. This $g-r$ minimum matches with the main peak of the light curves. From 43 to 100 days, the SN  slowly becomes redder again. A similar but less intense colour evolution is observed in $r-i$ and $i-z$. 

\subsection{Bolometric luminosity}
\label{sec:bolo}

Using the $groiz$ photometry, we build a pseudo-bolometric light curve for SN~2019cad. First, we interpolated the observed magnitudes using GPs to have the $groiz$ light curves with similar coverage at the same epochs. Then, extinction corrected (only galactic extinction) $groiz$ magnitudes were converted into fluxes at the effective wavelength of the corresponding filters. Next, we integrated a spectral energy distribution (SED) over the wavelengths covered, assuming zero flux beyond the integration limits. Fluxes were converted to luminosity using the distance adopted in Section~\ref{galaxy}. To estimate the full bolometric light curve, we extrapolated the SED constructed from the $groiz$. In the UV, we did a linear extrapolation to zero flux at 3000 \AA.
On the NIR side, the fluxes were extrapolated using a blackbody fit to the SED. This way to extrapolate to UV and NIR has been extensively used in the literature \citep[e.g.,][]{Folatelli06}. The full bolometric light curve of SN~2019cad for $E(B-V)_{Host}=0.0$ mag (red crosses) and $E(B-V)_{Host}=0.49$ mag (dashed line) are presented in the bottom panel of Figure~\ref{lc}. As seen, the limited photometric coverage at early and late times gives us larger uncertainties during the initial peak  as well as a slightly increase in luminosity after 100 days.

To estimate the luminosity of the initial and main peak, together with the slope of the tail, we use GPs. For the bolometric light curve assuming $E(B-V)=0.0$ mag, we find a main peak luminosity of $L_\text{bol}=5.1\times10^{42}$ erg s$^{-1}$ occurring at 43.6 days. The initial peak happens at 23 days at a luminosity of $1.7\times10^{41}$ erg s$^{-1}$. After $\sim60$ days, the  bolometric light curve declines at a rate of 5.75 mag per 100 days. These values are much higher than that expected from the \cofs\ decay, but they are
comparable to those obtained for the redder filters in the optical light curves.

\subsection{Spectral evolution}
\label{sec:spec}

\begin{figure*}
\centering
\includegraphics[width=0.7\textwidth]{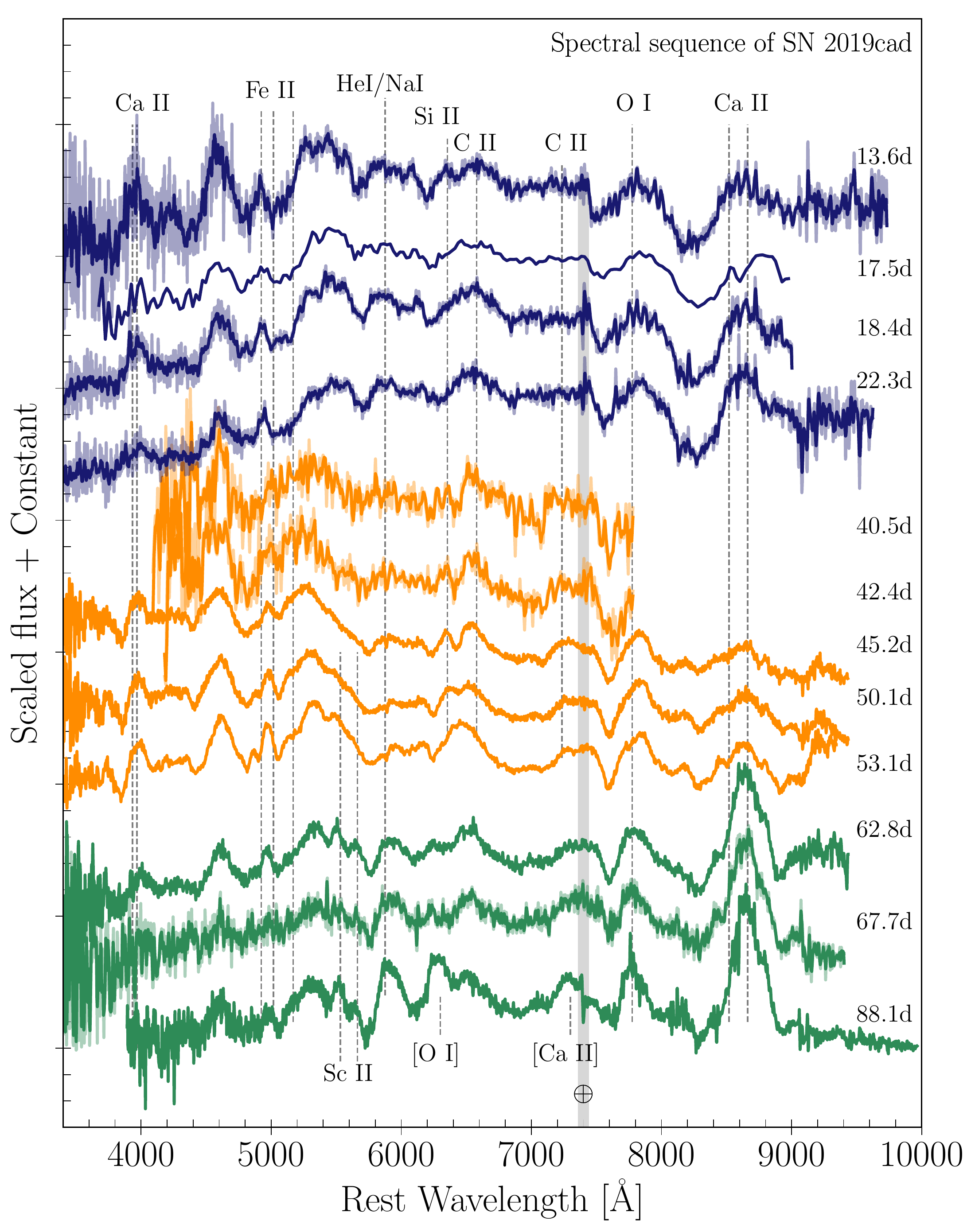}
\caption{Spectral sequence of SN~2019cad from 13.6 to 88.1 days from explosion. The phases are labeled on the right. Each spectrum has been corrected for Milky Way reddening and shifted vertically by an arbitrary amount for presentation. Vertical dashed lines indicate the rest position of the strongest lines. The colour of the spectra represents the different phases where they were observed: initial peak in blue, main peak in orange, and linear tail in green.}
\label{spectra}
\end{figure*}

Figure~\ref{spectra} shows the optical spectra of SN~2019cad from 13.6 days to 88.1 days past explosion. The first four spectra obtained during the initial peak (before 30 days) display the typical features of a SN~Ic (identifications confirmed by the \textsc{synow} \citep{Fisher00} fits; see below). At this early phase, the spectra are dominated by strong \ion{O}{i} $\lambda7774$, \ion{Ca}{ii} (H\&K and NIR triplet) and \ion{Fe}{ii} lines (around 5000 \AA). \ion{Si}{ii} $\lambda6355$ and \ion{Na}{i}\,D are also visible, together with a weak absorption of \ion{C}{ii} $\lambda6580$ and $\lambda7235$. From 13.6 to 22.3 days, the bluer part of the spectra diminishes due to \ion{Fe}{ii} line-blanketing,  \ion{Si}{ii} $\lambda6355$ becomes weak and \ion{C}{ii} $\lambda6580$ completely disappears. 

From 40 days past explosion, the spectra of SN~2019cad show a remarkable transformation, which coincides with the main light-curve peak. 
At 45 days, the spectrum is anew characterised by a blue continuum with weak features of \ion{Ca}{ii} (H\&K and NIR triplet), \ion{Na}{i}\,D, and \ion{Fe}{ii}. \ion{O}{i} $\lambda7774$ is now the strongest feature in the spectrum. \ion{Si}{ii} $\lambda6355$, \ion{C}{ii} $\lambda6580$ and $\lambda7235$ are visible again after almost complete disappearance at 22.3 days. At 50 days, \ion{C}{ii} $\lambda6580$ and $\lambda7235$ become weaker and are undetectable at 53 days. In the same period, the bluer part of the spectra shows three absorption features that correspond to \ion{Fe}{ii} $\lambda\lambda\lambda4924$, 5018 and 5169 lines, together with a clear detection of \ion{Sc}{ii} $\lambda5531$.

Later on, from day 62.8, the \ion{Ca}{ii} NIR triplet starts to show an emission component, suggesting the start of the nebular phase. \ion{Sc}{ii} $\lambda5531$ becomes stronger, while \ion{Sc}{ii} $\lambda5663$ is clearly detected. On the other hand, \ion{Si}{ii} $\lambda6355$ vanishes. We observe a significant decrease of the flux in the bluer part of the spectra, which is mainly produced by the line-blanketing.
In the last observation, at 88.1 days, the spectrum shows a combination of absorption and emission lines. The redder part of the spectrum starts to be dominated by forbidden emission lines of \ion{[O}{i]} $\lambda\lambda6300$, 6364 and \ion{[Ca}{ii]} $\lambda\lambda7291,$ 7323, while the bluer part is still dominated by the iron absorption lines. The \ion{Na}{i}\,D line presents a strong residual absorption component. At this stage, the \ion{Ca}{ii} NIR triplet is the strongest feature in the spectrum.

\begin{figure}
\centering
\includegraphics[width=\columnwidth]{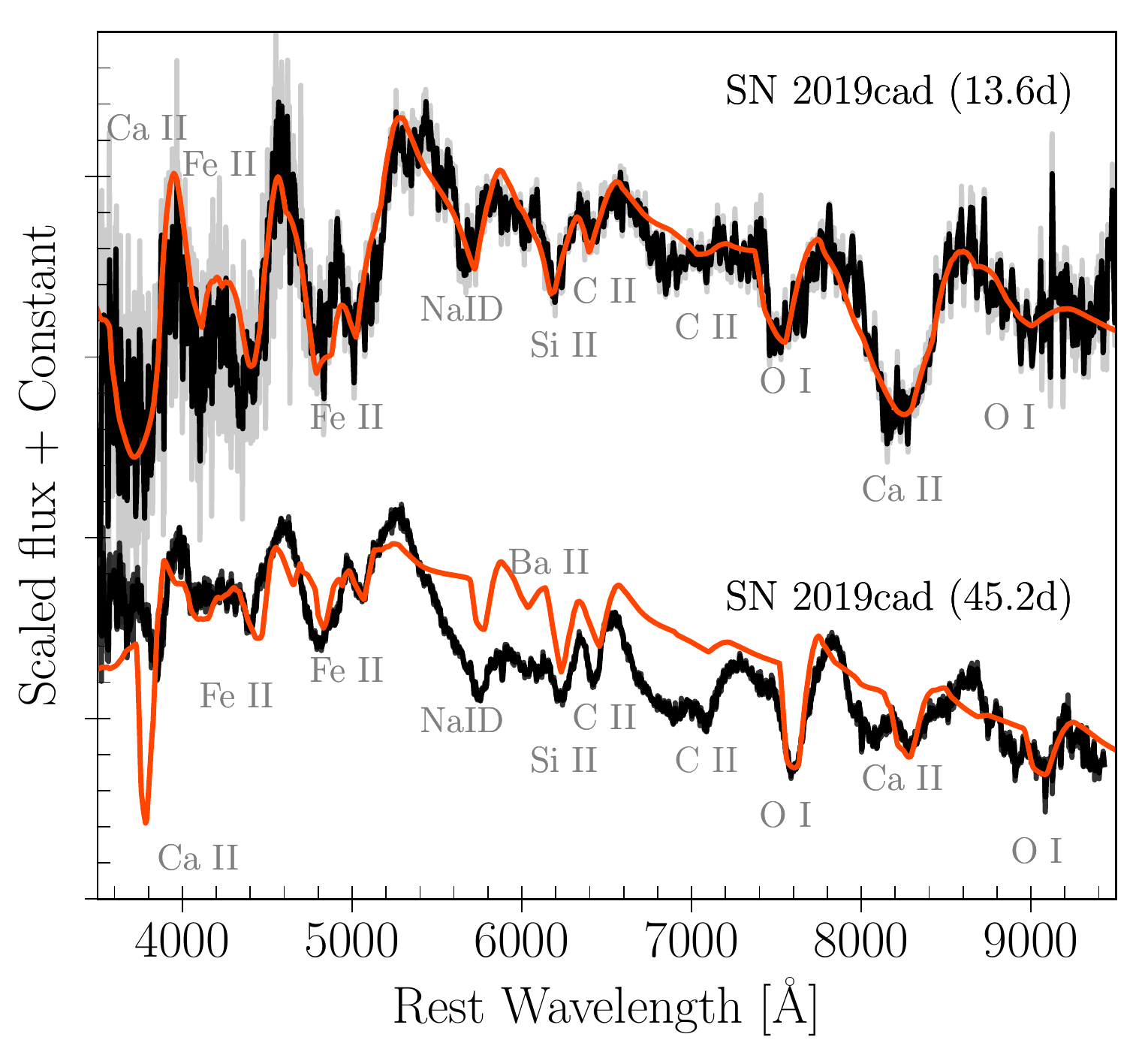}
\caption{Spectral comparison of SN~2019cad at 13.6 and 45.2 days and the \textsc{synow} fits. The \textsc{synow} synthetic spectra (orange) are overplotted on the observed spectra (black).}
\label{synow}
\end{figure}

To identify the lines in the spectra of SN~2019cad, we use the \textsc{synow} code \citep{Fisher00}. Figure~\ref{synow} shows the best fit obtained for the SN spectra at 13.6 and 45.2 days. For the first spectrum, we assume a blackbody temperature of $T_\text{bb}=5300$ K and a photospheric velocity of $v_\text{ph}=8600$ \kms, while for the spectrum at 45.2 days, we use  $T_\text{bb}=5600$ K and $v_\text{ph}=5500$ \kms. The synthetic spectrum at 13.6 days was obtained including \ion{Ca}{ii}, \ion{O}{i}, \ion{Na}{i}\,D, \ion{Si}{ii}, \ion{Fe}{ii} and \ion{C}{II}. Overall, the synthetic spectrum reproduces very well the observed features of SN~2019cad and helps us to confirm the two absorptions near $\sim6000$ \AA\ as \ion{Si}{ii} and \ion{C}{II}. For the second epoch, the synthetic spectrum  contains lines of \ion{Ca}{ii}, \ion{O}{i}, \ion{Na}{i}\,D, \ion{Si}{ii}, \ion{C}{II} and \ion{Ba}{ii}. Although we can reproduce most of the lines observed in SN~2019cad, we do not find a great match for the \ion{Ca}{ii} H \& K neither for the continuum in the region between 5500 and 7500 \AA. The difficulty to reproduce this second spectrum could be caused by the energy source powering the main peak, which could break some of the assumptions of \textsc{synow}. One interesting characteristic of these two spectra and their respective fits is the strength of the \ion{Si}{ii} and \ion{C}{ii} lines. Normally, these lines become weaker over time due to a temperature dependence: their strength decreases as the temperature decreases. However, for SN~2019cad the temperature rises at $\sim40$ days, and thus the \ion{Si}{ii} and \ion{C}{ii} are detected again. 
The presence of \ion{C}{ii} in the SN spectra could indicate a significant amount of carbon in the progenitor star.

\subsection{Expansion velocities}
\label{sec:vel}

\begin{figure}
\centering
\includegraphics[width=\columnwidth]{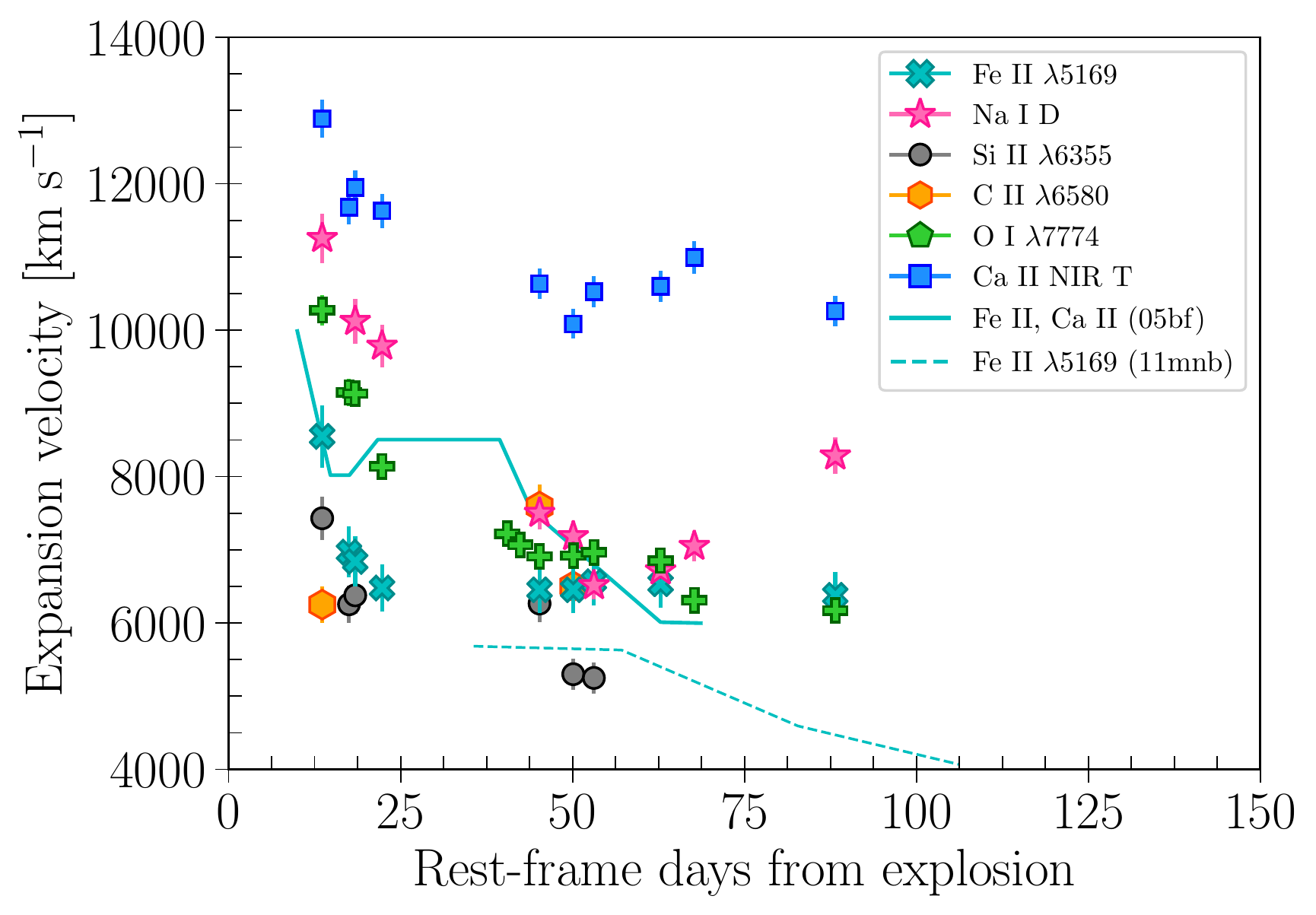}
\caption{Evolution of expansion velocities of SN~2019cad derived from the minimum of the absorption line. Epochs are with respect to the explosion. For comparison, we also include the \ion{Fe}{ii} and \ion{Ca}{ii} velocities of SN~2005bf \citep{Folatelli06} and \ion{Fe}{ii} $\lambda5169$ velocities of PTF11mnb \citep{Taddia18a}.
}
\label{expvel}
\end{figure}

We measure the expansion velocities of the ejecta from the minimum absorption flux of six spectral features. The velocity evolution of \ion{Ca}{ii} $\lambda\lambda8498$, 8542, 8662 triplet feature, \ion{Na}{i}\,D $\lambda5893$ \ion{O}{i} $\lambda7774$, \ion{Fe}{ii} $\lambda5169$, \ion{Si}{ii} $\lambda6355$ and \ion{C}{ii} $\lambda6580$ is presented in Figure~\ref{expvel}. At earlier phases (before 30 days), all lines have decreasing velocities, as expected for a homologous expansion. After 40 days, the velocities of \ion{Ca}{ii} and \ion{Na}{i}\,D slightly increase, while for the rest of the lines, they remain nearly constant. The change in the velocity evolution matches with the remarkable transformation observed in the light curves and suggests an additional source of energy.
The \ion{Ca}{ii} NIR triplet has the highest velocity, evolving from $\sim13000$ to $\sim10200$ \kms, while the \ion{Si}{ii} has the lowest one, decreasing from $\sim7400$ to $\sim5200$ \kms. This behaviour implies that the \ion{Ca}{ii} lines mainly form in the outer part of the ejecta and \ion{Si}{ii} lines form in deeper layers. \ion{Na}{i}\,D, \ion{Fe}{ii} and \ion{O}{i} have intermediate velocities (between $\sim11000$ and $\sim6000$ \kms). At early phases, \ion{Na}{i}\,D expands faster, followed by \ion{O}{i} and  \ion{Fe}{ii}, but after 40 days, all of them have similar velocities ($\sim7000$ \kms).
In general, the velocity range of SN~2019cad is comparable with other normal SN~Ic, however, the flat/rising velocity behaviour measured after 40 days is similar to that observed in SN~2005bf \citep{Tominaga05,Folatelli06} and PTF11mnb \citep{Taddia18a}. 

\subsection{Comparison with other SNe}
\label{sec:comp}

\begin{figure}
\centering
\includegraphics[width=\columnwidth]{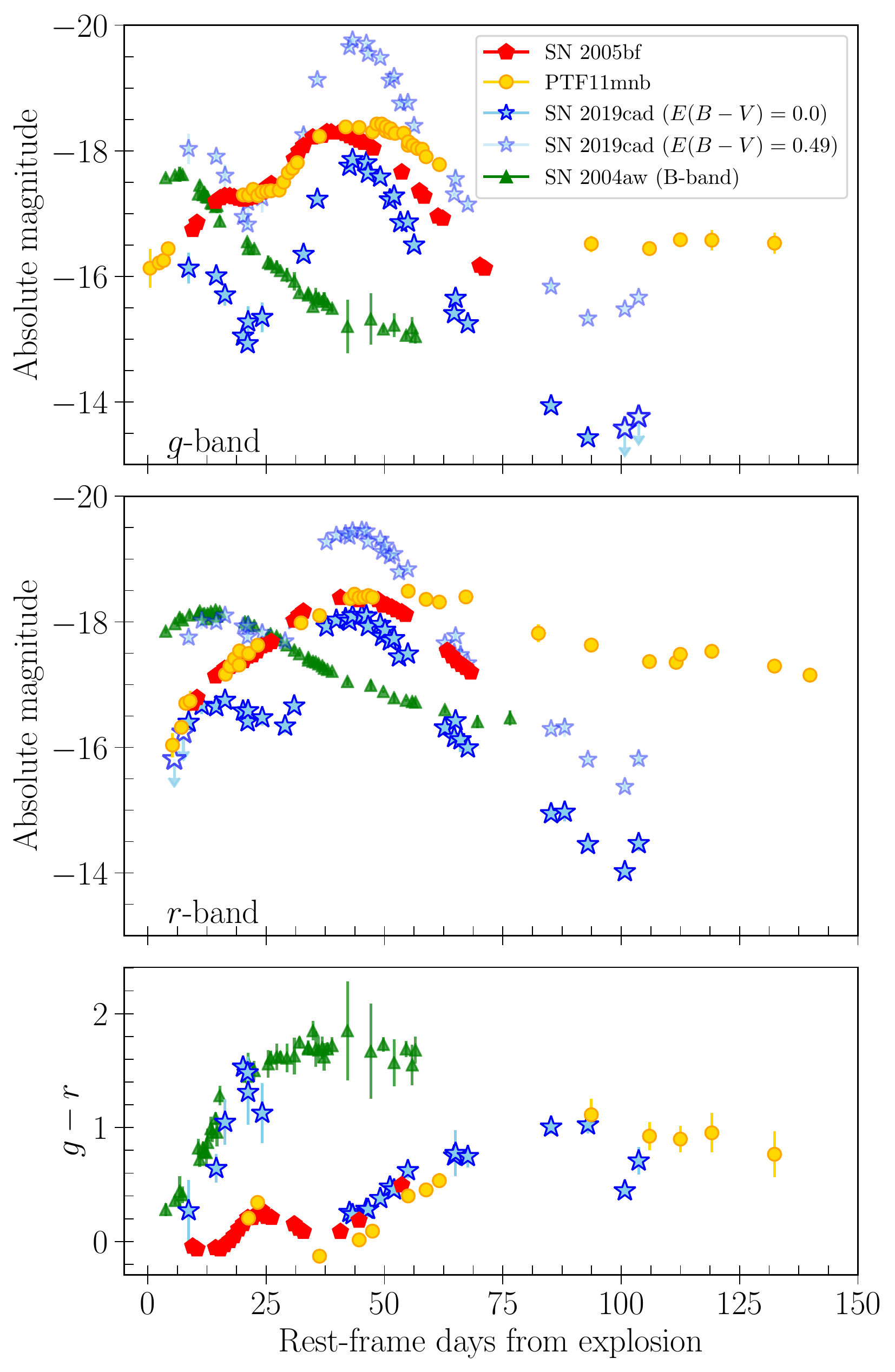}
\caption{Light curve and colour comparisons. \textbf{Upper and middle:} Comparison of the $g$-band (upper) and $r$-band (middle) light curves of SN~2019cad (blue stars) with the double-peaked light curves of SN~2005bf (red pentagons) and PTF11mnb (yellow circles), and the normal type Ic SN~2004aw (green triangles). Please note that for SN~2004aw we use the $B$ and $R$ bands. A light curve of SN~2019cad assuming an $E(B-V)=0.49$ mag is also included for reference. 
\textbf{Bottom:} Colour ($g-r$) curves of SN~2019cad and the comparison sample. Only Milky Way extinction corrections have been.
}
\label{LCcomp}
\end{figure}

\begin{figure}
\centering
\includegraphics[width=\columnwidth]{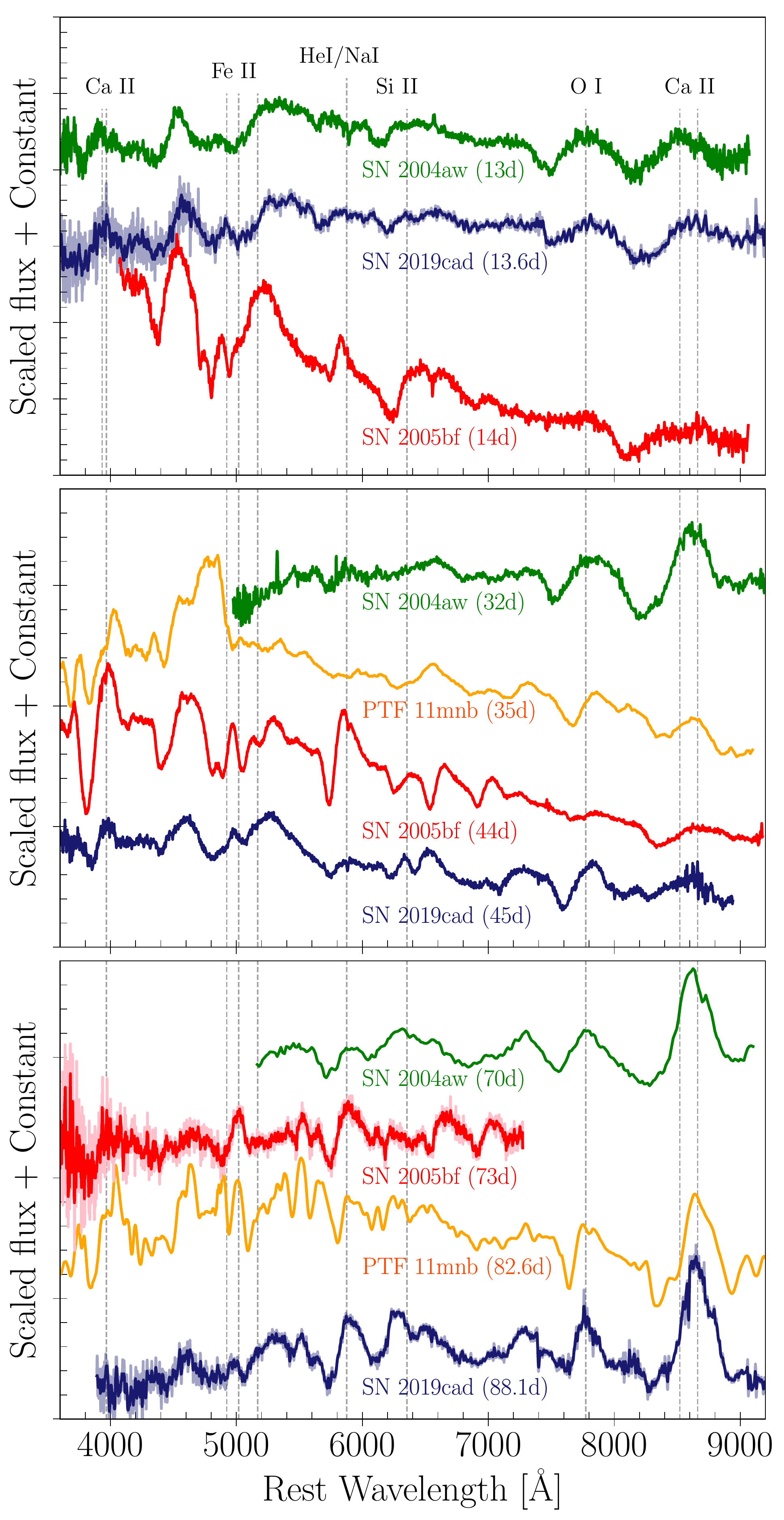}
\caption{Spectral comparison of SN~2019cad with SN~2005bf, PTF11mnb and SN~2004aw at three different epochs (13, 40 and 80 days).
}
\label{Speccomp}
\end{figure}

The photometric and spectroscopic evolution of SN~2019cad is unprecedented, however, its double-peaked light curves resemble the type Ib/c SN~2005bf \citep{Folatelli06} and the type Ic PTF11mnb \citep{Taddia18a}. From the spectral analysis, we find that SN~2019cad has the characteristic lines of a SN~Ic at early times, but around the main peak, the spectra show an unexpected evolution, characterised by a blue continuum and the presence of \ion{C}{ii} and \ion{Si}{ii} lines, which had disappeared at $\sim20$ days (Section~\ref{sec:spec}). Looking for objects with similar characteristics, we use \textsc{snid} \citep{Blondin07}. At early phase (before 30 days), we find a good match with the normal type Ic SN~2004aw \citep{Taubenberger06}, however, after the main peak, good spectral matches were not obtained. Based on the common features that SN~2019cad share with SN~2005bf, PTF11mnb (photometrically) and SN~2004aw (spectroscopically), we compare their light curves and spectra in Figures~\ref{LCcomp} and \ref{Speccomp}, respectively. Table~\ref{Comp_param} shows the main parameters measured for the double-peaked light curve SNe and SN~2004aw.

\begin{table*}
\begin{center}
\caption{Properties of SN~2019cad and the comparison sample}
\begin{tabular}{lcccccc}
\hline
Parameters                   & SN~2019cad$^{1}$ & SN~2005bf$^{2}$ & PTF11mnb$^{3}$           & SN~2004aw$^{4}$ \\
\hline
\hline
Host Galaxy                  &  UGC 4798        & MCG +00-27-5    & SDSS J003413.34+024832.9 & NGC 3997        \\
Explosion date (MJD)         & $58550.44\pm4$   & $53458.00$      & $55804.34\pm0.5$         & 53080.9         \\
Distance modulus             &  35.38           & 34.57           &  37.14                   &  34.26          \\
Spectroscopic classification &  Ic              &    Ib/c         &   Ic                     &  Ic             \\
$E(B-V)_{MW}$ (mag)          &  0.015           &    0.045        &   0.016                  &  0.20           \\
$E(B-V)_{Host}$ (mag)        &  0.0             &    0.0          &   0.0                    &  0.35           \\
\hline
\hline
\multicolumn{4}{c}{\textbf{$g$-band}}\\
Initial peak magnitude (mag) &$-16.35^{\star}$&   $-17.24$       &  $-17.39$                &   \nodata        \\
Initial peak epoch (d)       &     8          &      16.2        &    21.3                  &   \nodata        \\
Main peak magnitude (mag)    &$-17.83^{\star}$&   $-18.25$       &  $-18.37$                &$-17.63^{\dagger}$\\
Main peak epoch (d)          &   43.9         &      40          &    46.3                  &   7              \\
\hline
\hline
\multicolumn{4}{c}{\textbf{$r$-band}}\\
Initial peak magnitude (mag) &$-16.68^{\star}$&    \nodata       &   \nodata                &    \nodata       \\
Initial peak epoch (d)       &     15         &    \nodata       &   \nodata                &    \nodata       \\
Main peak magnitude (mag)    &$-18.10^{\star}$&    $-18.44$      &  $-18.45$                &$-18.14^{\dagger}$\\
Main peak epoch (d)          &    44.3        &        42        &    52.2                  &   14             \\
\hline
\end{tabular}
\begin{list}{}{}
\item \textbf{References:} 
\item $^{1}$ This work; $^{2}$ \citet{Folatelli06}; $^{3}$ \citet{Taddia18a}; $^{4}$ \citet{Taubenberger06}. \\
$^{\star}$ Absolute magnitudes assuming $E(B-V)_{Host}=0$ mag.\\
$^{\dagger}$ For SN~2004aw, we report the $B$ (instead $g$) and $R$ magnitudes, respectively.
\end{list}
\label{Comp_param}
\end{center}
\end{table*}

To compare the photometric evolution of these objects in the $g$ and $r$ bands, we present the light curves of SN~2019cad assuming an $E(B-V)=0.0$ mag, but we also correct the light curve for an $E(B-V)=0.49$ mag (clear blue stars). Additionally, we include the normal type Ic SN~2004aw (in $B$ and $R$). 
From the light curve comparison (Figure~\ref{LCcomp}, upper and middle panel), we find that SN~2019cad has a very pronounced initial peak, both in the $g$ and $r$-bands. While in SN~2005bf and PTF11mnb the initial peak decreases $\sim0.5$ mag in the $g$, in  SN~2019cad the brightness changes more than 1 mag. In the $r$ band, only SN~2019cad shows an initial peak. Instead, SN~2005bf and PTF11mnb present a shoulder, which is more conspicuous in PTF11mnb. 

Followed by the first maximum, a rise to the main peak is observed in all three objects. Whereas the brightness in SN~2005bf and PTF11mnb increase $\sim1$ mag in $g$, in SN~2019cad the rise is around 3 mag in $g$ and about 1.8 mag in $r$. For SN~2005bf, the main peak occurs at $\sim40$ days in $g$ and $\sim42$ days in $r$. PTF11mnb reaches it at $\sim46$ and $\sim52$ days, and SN~2019cad at 43.9 and 44.3 days, respectively.
SN~2019cad has a main peak that is the most narrow of those shown by these three double-peaked objects. At around 100 days, SN~2005bf has declined by more than 2 mag from peak in $r$-band, PTF11mnb only $\sim1$ mag and SN~2019cad more than 4 mag. 

On the other hand, SN~2004aw shows the typical light curve of a SN~Ic, with a main $R$-peak at $\sim14$ days from the explosion. In terms of luminosity, SN~2004aw is brighter than the initial peak of SN~2019cad (for $E(B-V)=0.0$ mag), but it is similar in the $r$-band when an $E(B-V)=0.49$ mag is assumed for SN~2019cad. 
Comparing the luminosity of the double-peaked objects, we find that SN~2019cad is always fainter than SN~2005bf and PTF11mnb in both the initial and main peak for $E(B-V)=0.0$ mag. However, if an $E(B-V)=0.49$ mag is assumed, SN~2019cad is $>1.5$ mag brighter than these two objects.   

From the colour curves (Figure~\ref{LCcomp}, bottom panel), SN~2019cad and SN~2004aw show a similar evolution during the first 20 days, being SN~2004aw slightly redder. In the same period, SN~2005bf and PTF11mnb show colours much bluer than SN~2019cad and SN~2004aw. This behaviour suggests that at early phases SN~2019cad was comparable to any other normal SN~Ic, however, after 20 days, the evolution of SN~2019cad changes radically: it becomes rapidly bluer reaching a minimum at $\sim44$ days (corresponding to the main peak in the light curve). From this epoch, the evolution and colours of SN~2019cad, SN~2005bf and PTF11mnb are similar. 

Figure~\ref{Speccomp} presents the spectral comparison of SN~2019cad with SN~2005bf, PTF11mnb and SN~2004aw at three different epochs: $\sim13$, $\sim40$ and $\sim80$ days from explosion. At around 13 days, SN~2019cad is very similar to the normal type Ic SN~2004aw. They show strong lines of \ion{Ca}{ii}, \ion{O}{i} and \ion{Fe}{ii}. The spectrum of SN~2005bf has a similar \ion{Ca}{ii} absorption line to SN~2019cad, however, the rest of the spectrum has significant differences, such as, weak detection of \ion{He}{i} lines, lack of \ion{O}{i} and a strong line around $\sim6265$ \AA, which has been identified as a high velocity feature of H$\alpha$ \citep{Folatelli06}. These properties could indicate that SN~2019cad is similar, but more stripped than SN~2005bf.
By $\sim40$ days, all objects in the comparison sample present very different spectra. Both SN~2019cad and SN~2004aw have a prominent \ion{O}{i} line, however, the \ion{Ca}{ii} absorption feature is shallow in SN~2019cad. For SN~2005bf, the \ion{He}{i} lines are exposed, but they are missing in the rest of the objects. SN~2005bf, PTF11mnb and SN~2019cad have blue spectra compared to SN~2004aw. At this phase, SN~2019cad has prominent lines of \ion{C}{ii} and \ion{Si}{ii}, which are not clearly seen in the other objects. At $\sim80$ days, most of the objects show some signatures of emission lines, which indicate the beginning of the nebular phase. Although SN~2019cad, SN~2005bf and PTF11mnb have similar light curve morphology, and comparable colours after 40 days, their spectra are completely different. This spectral diversity could suggest differences in their progenitor stars.  

When we compared the evolution of the \ion{Fe}{ii} velocities of SN~2019cad and those from the double-peaked SN~2005bf and PTF11mnb, we notice that SN~2019cad lies between these two objects (see Figure~\ref{expvel}). Before 15 days, SN~2019cad and SN~2005bf have comparable velocities, but after that, the velocity of SN~2019cad continues decreasing, while the velocity of SN~2005bf has a slight rise, and then it is flat. PTF11mnb, the object with the lowest velocities, has a flat evolution between 38 and 60 days. A flat evolution is also observed in SN~2019cad later than 40 days.

\section{Light curve modelling}
\label{sec:model}
In order to understand the sources that could be responsible for enhancing the light curve of SN~2019cad, we have explored the main competing ideas that were presented in relation to peculiar double-peaked SNe, SN~2005bf \citep{Folatelli06} and PTF11mnb \citep{Taddia18a}. Like SN~2019cad, these objects present a rise and decline in luminosity prior to the main peak.
Such early behavior cannot be reconciled with an extended envelope as used to model cooling emission in SNe~IIb or by assuming the presence of a small amount of circumstellar material \citep{Bersten12,Nakar14}. We performed a brief exploration of these possibilities with no success. The main reason was the duration of the SN~2019cad first peak. Assuming the presence of a thin envelope, we can only reproduce a first peak with a duration of less than 10 days and a decreasing luminosity due to the cooling process. As a result of this cooling process and the existence of an additional source that heats the ejecta, usually radioactivity, the light curve shows a minimum. Similar outcome can be obtained assuming the presence of a local and low mass CSM (see also \citet{Jin21} for comparable results). Nevertheless, more extreme CSM conditions, as expected to explain hydrogen-rich SNe with narrow emission lines (SN~IIn) or some superluminous SNe (SLSNe), maybe could explain the early and/or main peak in the SN light curve. However, we did not explore these possibilities here.
Instead, we analysed in detail the following scenarios: (1) a double nickel distribution (Section \ref{sec:doubleniquel}) and (2) the first peak powered by nickel but the second one by a magnetar (Section \ref{sec:magnetar}). In this analysis, we have assumed a zero host extinction (see discussion in Section~\ref{sec:extinction}).

\subsection{Double \nifs\ distribution}
\label{sec:doubleniquel}

\begin{figure}
\centering
\includegraphics[width=\columnwidth]{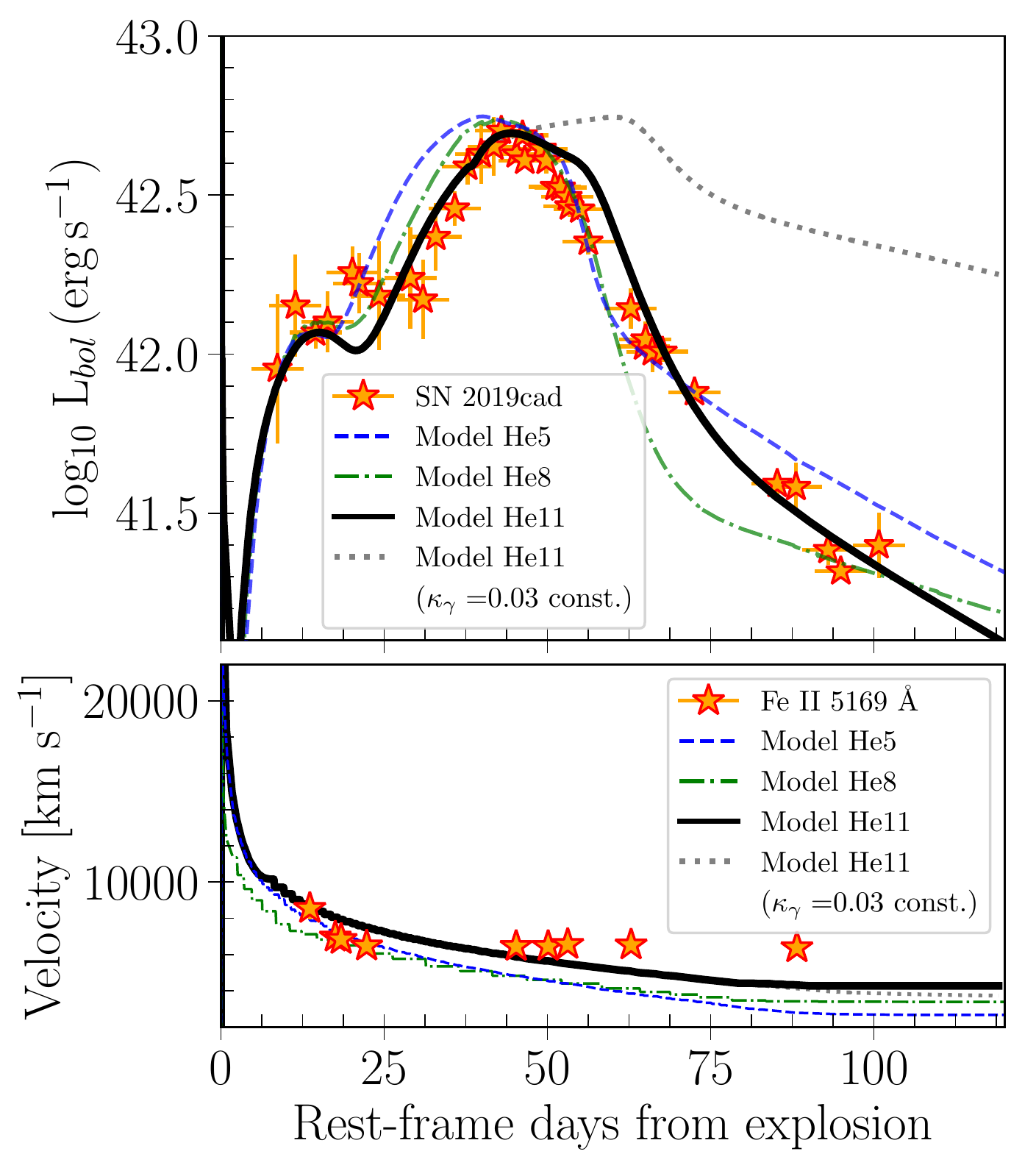}
\caption{\textbf{Top:} Comparison between our preferred double-peaked \nifs\ model (black solid line) and the SN observations (orange stars). For this model, we use a progenitor with a pre-SN mass of 11 \Msun\ (He11), corresponding to a Zero Age Main Sequence mass of 30 \Msun, an explosion energy of $3.5\times10^{51}$ erg and ejected mass of 9.5 \Msun. For comparison, we also include the double \nifs\ distribution models for lower pre-SN masses: 5 (He5; in blue dashed line) and  8 (He8; green dash-dotted line). In all three cases, an enhancement of the gamma-ray leakage is assumed at around the date of $L$-main peak. Models with lower mass produces an earlier main peak and overestimate the rise time luminosity. 
The He11 model assuming a constant $\kappa_{\gamma}=0.03$ (dotted grey line) is also presented as reference.
\textbf{Bottom:} Evolution of the He11 (black solid line), He5 (in blue dashed line) and  He8 (green dash-dotted) line photospheric velocity compared with the \ion{Fe}{ii} velocity of SN~2019cad.
}
\label{model}
\end{figure}

SN~2019cad is luminous enough to envisage that outflows were involved in its explosion. Several studies (see e.g., \citealt{MacFadyen2001, Banerjee13}) have considered SNe with outflows or jets in relation to gamma-ray burst. Jets can induce nucleosynthesis of radioactive elements at the outer layers of the ejecta before the shock front of the SN arrives  \citep{Nishimura15}. The outer nickel produced in this way could be responsible of the first peak observed in the light curve of SN 2019cad as explored for SN 2005bf \citep{Tominaga05,Folatelli06}, PTF11mnb \citep{Taddia18a} and SN~2008D \citep{Bersten13}.
The lack of high-energy emission could be explained by the inability of the jets to breakout the SN surface, or by geometric reasons. Although we did not find any high-emission reported, we cannot rule out their possible existence.

To analyse the double \nifs\ possibility, we performed hydrodynamic calculations using different helium rich progenitors as hydrostatic initial conditions. Specifically, we have tested models with masses of 5 (He5) and  8 \Msun\ (He8) at the pre-supernovae stage.
These models were calculated by \citet{Nomoto88} and correspond to main sequence stars of 18, and 25 \Msun, respectively. A more massive progenitor of 11 \Msun\ (He11) corresponding to a main sequence star of 30 \Msun\ computed with \textsc{mesa} code \citep{Paxton11} was also used. We have artificially exploded these configurations using the code presented in \citet{Bersten11}; a one-dimensional radiation hydrodynamical code which assumed flux-limited diffusion approximation and  gray transfer for gamma photons produced during the radioactive decay. The code assumes that the positrons are fully trapped all times and fully deposit their energy before annihilation following \citet{Sutherland84}.   
After several calculations, we found that the highest mass progenitor (He11 in our case) is favored to explain the double-peaked light curve of SN~2019cad (see Figure~\ref{model}). This is because it allows the large temporal departure between the light curve peaks as a consequence of depositing the inner and outer nickel at a larger distance in mass coordinate.
For lower mass progenitors, even if the general luminosity trend in the light curve can be reached, the resulting $L_{\rm bol}$ between 20--40 days is far above that observed, with the morphology of a minimum around 25 days poorly reproduced. 

\begin{figure}
\centering
\includegraphics[width=\columnwidth]{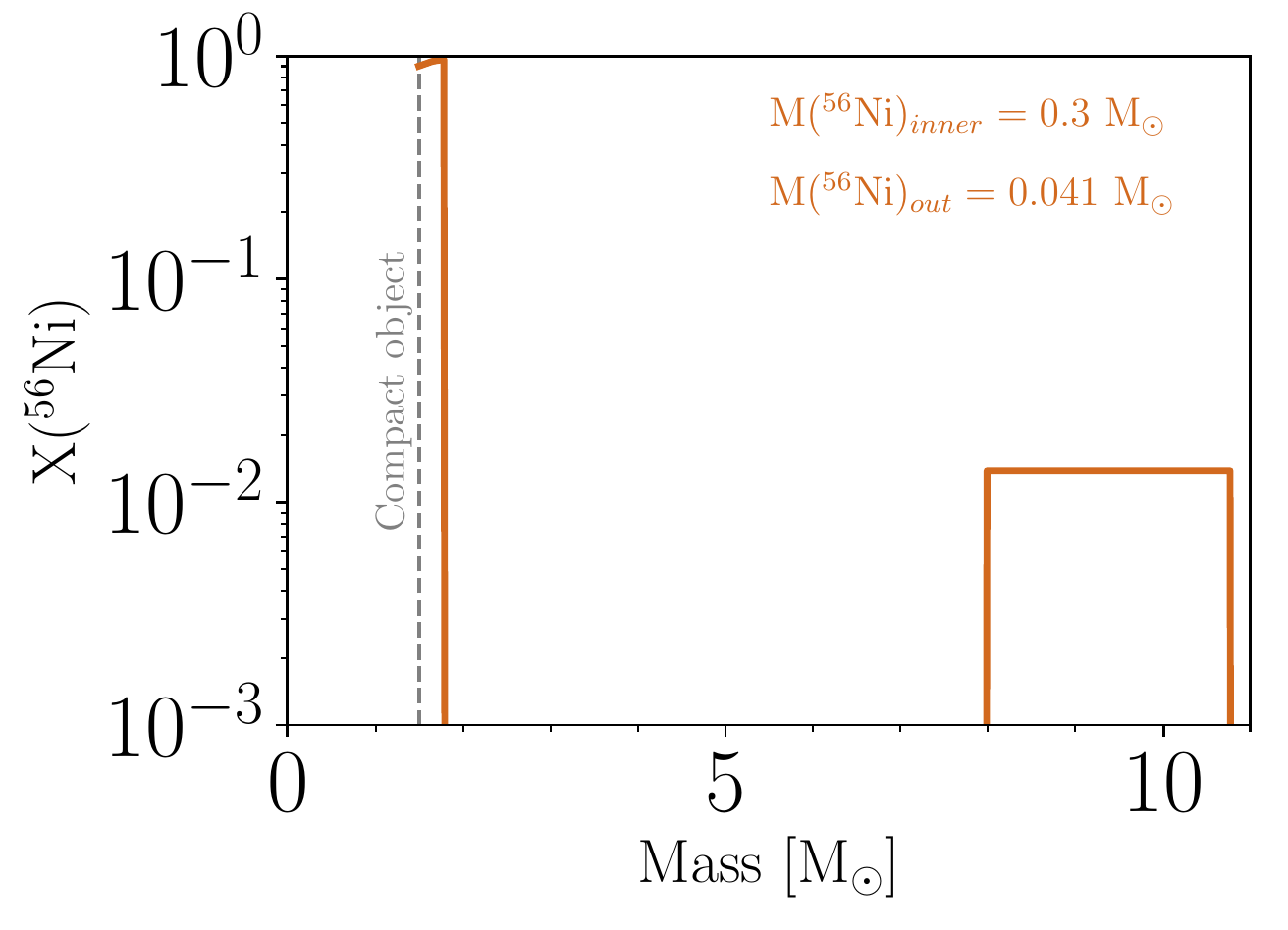}
\caption{The \nifs\ abundance profile of our preferred model powered by the double-peaked nickel distribution. The stellar surface is at 500 R$_{\odot}$ for $\sim11$ M$_{\odot}$ before explosion (i.e. the outer component is not extended up to the surface). Note that the inner component shows a slightly decline due to presence of other heavy elements like Si, Ar or Ca.
}
\label{perfil_Ni}
\end{figure}

Our best case was produced by He11 assuming an explosion energy of $E_{\rm exp}=3.5\, \times 10^{51}$ erg, an ejected mass of $M_{\rm ej}=$ 9.5~\Msun\ and the formation of a neutron star of $\approx 1.5$ \Msun. Figure~\ref{model} shows this model compared with the observations of SN 2019cad. The double-peaked morphology of the light curve was obtained assuming the nickel distribution presented in Figure~\ref{perfil_Ni}.
We have had to consider a concentrated inner component close to the compact remnant ($1.5$ \Msun), similar to what was found for SN~2005bf \citep{Folatelli06}, plus an external \nifs\ component. 
In the model presented, the external nickel needs to be close but below the surface to fit the time scale of the first peak of the light curve. We found that an external \nifs\  mass of $0.041$ \Msun\ satisfied our requirements. This represents a small fraction of the inner component, with a mass of $0.3$ \Msun.

Other \nifs\ distributions have been also explored, in particular, we analysed if a smoother transition between the inner and outer \nifs\ component could improve the light curve modelling, but we did not find substantial differences with the distribution showed in Figure~\ref{perfil_Ni}. Therefore, we preferred to set a simplistic case with an inner \nifs-rich layer (the abundance mimics the one of Si-Ar-Ca, i.e. increasing outward) and an outer component with constant nickel abundance. These are well departed zones, with no nickel in the middle (see Figure~\ref{perfil_Ni}).

In addition to the early peculiar morphology, SN~2019cad shows a steeper decline in the light curve after the main peak, similar to that observed in SN~2005bf. Assuming that this peak was powered by nickel, an artificial reduction in the gamma-ray trapping was proposed to explain the post-peak behaviour of SN ~2005bf \citep{Tominaga05,Folatelli06}. An equivalent treatment, as an ad-hoc leakage factor used by \cite{2017vreeswijk} also resembles the incomplete trapping of gamma-rays in \cite{Drout2013}. Following this idea, we have modified the gamma-ray opacity from $\kappa_\gamma=0.03$ to $\kappa_\gamma=0.0005$ cm$^2$ g$^{-1}$ around the mean light curve peak of SN~2019cad. 
By default our code uses a fixed gamma-ray opacity, however this can be changed as required. The values mentioned above implies a decrease of the opacity by a factor of 60, which in turn allows a much larger gamma-ray leakage. We cannot provide a thorough physical justification for this assumption as it was made to fit the rapid light curve post-maximum decline. However, such a reduction of the gamma-ray opacity might be related to the asymmetry of the explosion that produces extremely inhomogeneous ejecta (see also e.g., \citealt{Maeda07}).

The model presented reproduces the light-curve properties of SN~2019cad reasonably well: the double-peaked morphology and the fast decline after the peak. However, some discrepancies appear around $\approx$ 25 days where the observations show a slightly higher luminosity than the model. 
Finally, we note that, although we have not considered the photospheric velocity evolution in the modelling (which could lead to a possible degeneracy in the parameters found, see discussion in \citealt{Martinez20} and references therein), the model gives a relatively good representation of the \ion{Fe}{ii} $\lambda5169$ velocities as shown in the bottom panel of Figure~\ref{model}.

While we consider an $E(B-V)_{Host}=0.0$ mag for our analysis, we also explore the effects of having an $E(B-V)_{Host}=0.49$ mag. As shown in the bottom panel of Figure~\ref{lc}, the main effect of assuming none negligible extinction is to increase the luminosity, and at first order, we can consider that the shape of the light curve remains similar. An increase in luminosity can be produced assuming a larger nickel mass. To reproduce a brighter light curve, our model requires an external nickel component of 0.132 \Msun\ and an internal component of 0.904 \Msun. These values are $\sim3$ times higher than our original model. Table~\ref{Mparam} summarises the values that better reproduce the light curve of SN~2019cad assuming an $E(B-V)_{Host}=0.0$ and 0.49 mag, respectively.

\begin{table}
\begin{center}
\scriptsize
\caption{SN~2019cad light curve modelling parameters}
\setlength{\tabcolsep}{4pt}
\begin{tabular}{cccccccc}
\hline
\hline
 &  $E(B-V)_{Host}=0.0$ mag & $E(B-V)_{Host}=0.49$ mag \\
\hline
\multicolumn{3}{c}{\textbf{Double \nifs\ distribution model}}\\
Internal \nifs\ component & 0.300 \Msun & 0.904 \Msun \\
External \nifs\ component & 0.041 \Msun & 0.132 \Msun \\
\hline
\multicolumn{3}{c}{\textbf{Double \nifs\ distribution plus magnetar model}}\\
Initial period (P) & 11 ms & 4 ms \\
Magnetic field (B) & $26\times 10^{14}$~G & $12~\times 10^{14}$~G & \\
External \nifs\ component & 0.041 \Msun & 0.132 \Msun \\
\hline
\hline
\end{tabular}
\label{Mparam}
\end{center}
\end{table}

\subsection{Magnetar}
\label{sec:magnetar}

\begin{figure}
\centering
\includegraphics[width=\columnwidth]{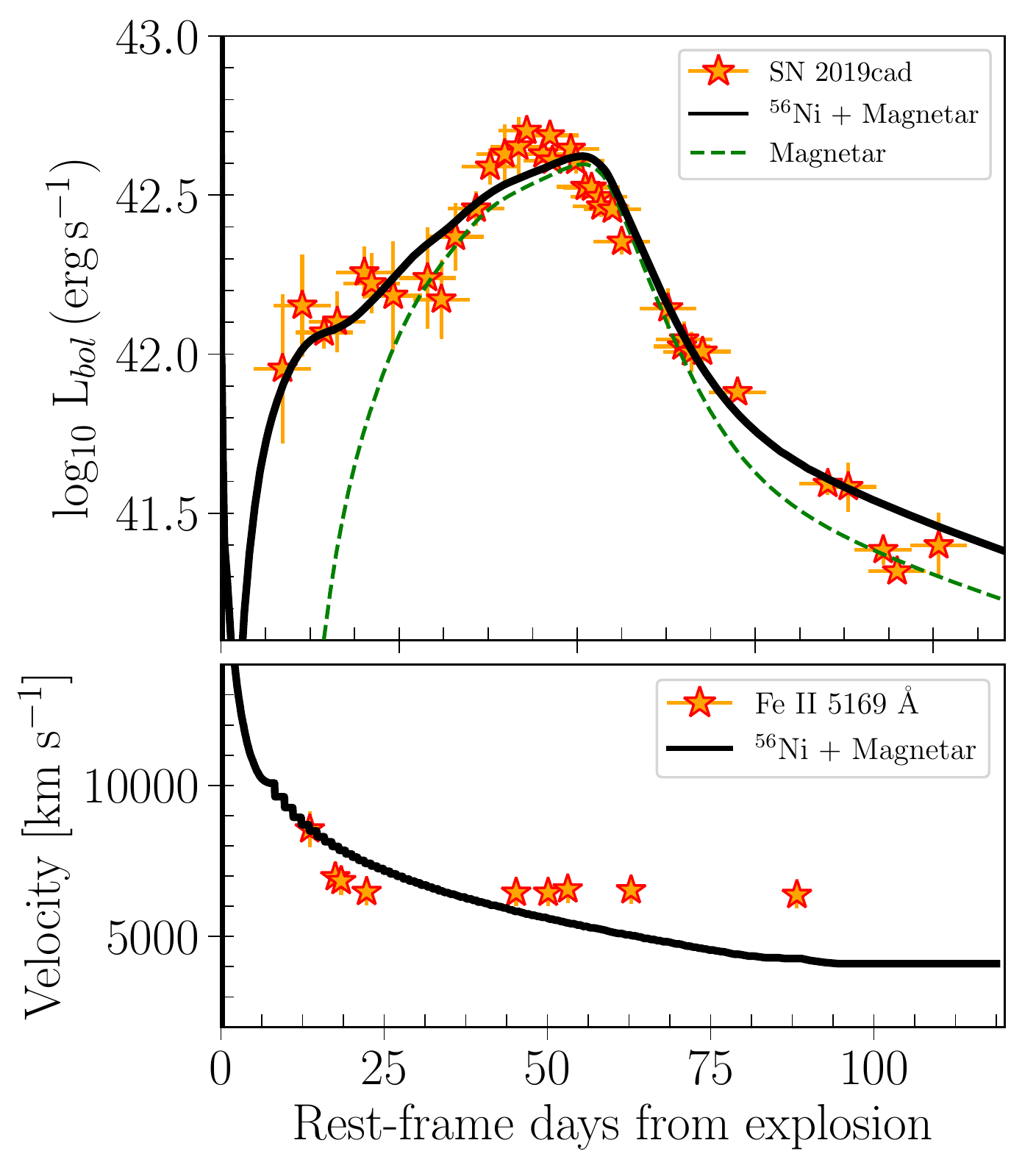}
\caption{\textbf{Top:} The magnetar plus nickel model for a progenitor mass of 11 \Msun\ (see Section~\ref{sec:magnetar}) compared with the light curve of SN~2019cad. For the first maximum, an outer \nifs\, mass of $0.041$~\Msun\ is assumed. 
The second, main peak was modeled with an inner mass of \nifs\ $0.05$~\Msun\ and a magnetar with an initial period $P=11$~ms  and a magnetic field of $B=26\times 10^{14}$~G (dashed line). 
The calculated light curve with both \nifs\ components and the magnetar is shown in black solid line. \textbf{Bottom:} Evolution of the magnetar plus nickel model photospheric velocity compared with the \ion{Fe}{ii} velocity of SN~2019cad.}
\label{magnetar_m}
\end{figure}

The idea that a strongly magnetised neutron star or magnetar could have been the powering source of the main peak in SN~2005bf was proposed by \citet{Maeda07}. Late time photometry of SN~2005bf has provided further support to this idea. 

The research on young magnetars into SNe has vastly grown in last years, with suggestions for regular \citep[see e.g.,][]{2017Sukhbold}, bright \citep{Inserra13a, Dessart18, Orellana18}, and peculiar SNe \citep{Bersten16,Woosley18}. Recent results provide new insights about the magnetar properties. Detailed simulations have shown that magnetars can accelerate iron-group elements deep in the ejecta, and explain the high-velocity Fe lines observed in some core-collapse SNe \citep{Chen2020}. Another interesting characteristic is that the magnetar creation might be accompanied by jets \citep[e.g.,][]{Soker16}. Based on this assumption, we propose that a magnetar and a double nickel distribution can, in principle, be combined. To explore this possibility for SN~2019cad, we applied a version of the hydrodynamical code that incorporates the extra power source of magnetar \citep{Bersten16} and assumes that this energy is deposited at the inner layers of the ejecta. 
For the low mass progenitors we did not find a promising case, therefore the He11 configuration was again preferred. We performed a wide exploration of the magnetar parameter space, for fixed standard properties of the neutron star, and assuming the spin-down proceeds through vacuum dipole radiation (braking index $n=3$). 
We obtained an acceptable match for the main peak for $P\sim 11$~ms and $B\sim 26\times 10^{14}$~G (see Figure \ref{magnetar_m}), with an explosion energy of  $E_{\rm exp}=3.5\times10^{51}$ erg and a conservative \nifs\ mass ($\sim 0.05$ \Msun) at the inner layers of the progenitor. This inner nickel mass is constrained by the observations at the tail of the light curve. We set the gamma-opacity to a usual value, $\kappa_{\gamma}=0.03$~cm$^2$~g$^{-1}$ during the complete evolution.
Note that the energy of the magnetar is assumed to be deposited at the inner layers of the ejecta as internal energy.
The magnetar model alone cannot reproduce the two peaks observed in the light curve of SN~2019cad. Therefore, we resume to the presence of some external nickel, and turn on the same configuration of the outer $\sim0.04$~\Msun\ \nifs\ depicted in Figure~\ref{perfil_Ni}.

In the hybrid model proposed for PTF11mnb \citep{Taddia18a}, the magnetar-powered light curve is explored through the  semianalytic prescription of \citet{Kasen10b} in combination with the diffusion one-zone model of \citet{Arnett82} for the \nifs\ radioactive decay. In order to obtain the maximum time scale with that treatment, the magnetar must ignite with a delay of several days from the explosion. We note the physical difficulty to explain such delay. The simulations of post-collapse are consistent with evolutionary timescales of seconds between the proto-neutron star birth and the moment when the stable neutron star holds the high angular momentum and strong magnetic field to initiates the spin-down phase \citep{2001Mezzacappa,  2007Metzger,2021Aloy}.
However, in our hydrodynamical implementation of the magnetar model assuming a massive (He11) progenitor, such delay was not needed and the magnetar was turned-on at the same time as the SN explosion, providing a more reliable prescription. In addition, this model did not require a reduction of the gamma-ray opacity.

Lastly, we explore the parameter space required to model a brighter light curve (assuming an $E(B-V)_{Host}=0.49$ mag, see Figure~\ref{lc}). As shown in Figure~\ref{magnetar_m}, in our double \nifs\ distribution plus magnetar model, the effect of the magnetar is more important for the main peak, while the impact of the \nifs\ mass is significant for the initial peak. It has been shown that in a magnetar, the peak luminosity is mostly affected by the initial period ($P$) and the magnetic field ($B$) \citep[e.g.,][]{Kasen10b,Dessart19}. Taking these three parameters ($P$, $B$ and \nifs\ mass) into account, we can reproduce the new light curve. To reach the main-peak luminosity, while preserving its timescale, a magnetar with $P\sim 4$~ms and $B\sim12~\times 10^{14}$~G is necessary. To reproduce the initial peak, we need to include an external nickel component of 0.132 \Msun. These parameters, summarised in Table~\ref{Mparam}, are within the range of previously published values (e.g., \citealt{Dessart19} for the magnetar and \citealt{Anderson19}, for the \nifs\ mass).

\section{Discussion and conclusions}
\label{sec:disc}

We have presented the spectroscopic and photometric analysis of the type Ic SN~2019cad during the first 100 days from explosion. Located at projected distance of 4.6 kpc from the centre of \ion{H}{i}-rich galaxy UGC~4798, SN~2019cad presents an uncommon light curve evolution, which resembles the peculiar type Ib/c SN~2005bf and the type Ic  PTF11mnb. That is, an initial peak at about 10-15 days from explosion followed  by a main peak at $\sim45$ days from explosion. While the luminosity of the first peak (${\rm M}_{r}= -16.68$ mag) is within the ranges of normal hydrogen-free events \citep[e.g.][]{Taddia18a}, the second one (${\rm M}^{max}_{r}= -18.10$ mag) is brighter than normal, but consistent with that observed in double-peaked SNe, if the extinction is $E(B-V)=0$ mag. From the spectroscopic point of view, SN~2019cad shows important differences with respect to SN~2005bf and PTF11mnb. 
During the first 30 days (first peak), the spectra of SN~2019cad are comparable with typical SNe~Ic, however, when the light curve goes up to the main peak (at around 45 days), the spectra display an unusual transformation, easily noticeable by the presence of the \ion{Si}{ii} and \ion{C}{ii} lines. The presence of these lines were confirmed by the \textsc{synow} fits. At these later epochs, no spectral matches were found, suggesting that SN~2019cad has an unique evolution.

To try to understand the peculiar photometric behaviour of SN~2019cad, we have explored two possible scenarios, both assuming a progenitor with a pre-SN mass of 11 \Msun, corresponding to a Zero Age Main Sequence mass of 30 \Msun, an explosion energy of 3.5 $\times 10^{51}$ erg and ejected mass of 9.5 \Msun, assuming the formation of a compact object with a 1.5 \Msun\ and a double-peaked \nifs\ distribution, with an outer, though below the surface, low-mass component of 0.04 \Msun. For the main peak we obtained
(1) an inner nickel component of 0.3 \Msun (section \ref{sec:doubleniquel}) ; or 
(2) a \nifs\ plus magnetar model (section \ref{sec:magnetar}). Specifically, the magnetar properties found are an initial rotation period $P\sim 11$ms and magnetic field strength of $B\sim 26\times 10^{14}$~G. A discussion of how the neutron star properties can affect the values of $P$ and $B$ can be found in \cite{Bersten16}. 
With this in mind, as well as other simplifications of the 1D model, the parameters obtained here for the magnetar should be considered approximate values.

While either model can reproduce the overall morphology observed in SN~2019cad, the double \nifs \, distribution model provided a bit better representation of the double-peaked light curve. However, to reproduce the late ($t>60$~d) behaviour
of fast decline post-maximum, we artificially reduced the gamma-ray trapping. Similar approaches were explored by \citet{Tominaga05,Folatelli06} to fit the decline observed in SN~2005bf, but this was not necessary in the case of SN PTF11mnb. For the magnetar model, an internal component of \nifs\ can be present, involving at most some 0.05 \Msun. The gamma opacity was not needed to be reduced.  The draw-back in this case is the worse fit of the bolometric light curve at the transition (around $t=25$~d) and up to the main peak. 

The amount of nickel found for the double \nifs\ distribution model is larger than the expected for hydrogen-free SNe \citep[see e.g.,][]{Prentice16,Taddia18,Anderson19}, while for the the model including the magnetar is within the range of these objects. The progenitor mass  (pre-SN mass of 11 \Msun, corresponding to a Zero Age Main Sequence mass of 30 \Msun) is also large, but comparable with that of SN~2005bf \citep{Tominaga05,Folatelli06,Maeda07}.
The outer \nifs\ component in the stellar interior that we assumed to be related with jets could be hinted by high-energy emission as in GRB-SNe explosions. 
Although, as this SN is quite peculiar, it may be a failed GRB \citep{Huang2002} i.e. having jet-like outflows
without sufficient Lorentz factor, or even a GRB with collimated jets that do not point in the direction of the observer. Archival searches for detections of SN2019cad at other wavelengths are therefore encouraged.

As discussed in Section~\ref{sec:extinction}, considerable discrepancies in the reddening estimation for SN~2019cad were found with different methods, and an $E(B-V)_{Host}=0.0$ mag was adopted for our analysis. However, based on the strong narrow \ion{Na}{i}\,D absorption feature detected in the spectra, we deduce that SN~2019cad may suffer a significant reddening. In section~\ref{sec:comp}, we compared SN~2019cad with similar objects and briefly discussed the implications of a higher reddening. By adopting an $E(B-V)_{Host}=0.49$ mag, we found that the light curve has a significant increase in brightness. To reach the new luminosity by using the double \nifs\ distribution model, an external nickel component of 0.132 \Msun\ and a value of 0.904 \Msun\ for the internal component were needed. We note that a \nifs\ mass of 0.904 \Msun\ is huge compared with the values measured in normal stripped-envelope SNe (mean value of 0.293 \Msun; \citealt{Anderson19}), and is beyond those estimated for even extreme objects such as GRB-SNe ($\sim0.3-0.6$ \Msun; e.g., \citealt{Cano17}). These gigantic values disfavour this model.
On the other hand, assuming a magnetar model, the luminosity could be easily reach by changing the initial period and the magnetic field. However, as our model requires nickel to reproduce the initial peak, an increase in the nickel mass is essential. In our preferred case, the external nickel component is, again, 0.132 \Msun, but $P\sim 4$~ms, which is a few times larger than the theoretical limit for break-up ($P\sim2$~ms; \citealt{Dessart19}). In order to maintain the time of the maximum in the light curve, the magnetic field should be $B\sim12~\times 10^{14}$~G. These values favour the double \nifs\ plus magnetar scenario.

Summarising, SN~2019cad is within a rare type of event, with only two previous examples, SN~2005bf and PTF11mnb. Despite showing some similar features, these three supernova do show key differences. The similarities in the light curve could suggest a similar explosion mechanism, while the diversity in the spectra could imply different progenitor evolution, e.g., distinctive grades of mass loss.
Although we have found two models that can explain the main photometric properties of SN~2019cad, there are still many issues in our understanding of these objects and the ultimate source of energy to power them is still a mystery.

\section*{Acknowledgements}

We thank the anonymous referee for the comments and suggestions that have helped to improve the paper.

We are grateful to Peter Jonker who enabled the WHT observation of this target during his program W19AN003.
We thank Peter Brown its contribution with data from the Neil Gehrels \textit{Swift} Observatory.

CPG and MS acknowledge support from EU/FP7-ERC grant No. [615929].
MO acknowledges support from UNRN PI2018 40B696 grant.
GP acknowledges support by ANID – Millennium Science Initiative – ICN12\_009. 
NER acknowledges support from MIUR, PRIN 2017 (grant 20179ZF5KS). 
MF is supported by a Royal Society - Science Foundation Ireland University Research Fellowship.
MS is supported by generous grants from VILLUM FONDEN (13261, 28021) and by a project grant (8021-00170B) from the Independent Research Fund Denmark.
LG was funded by the European Union's Horizon 2020 research and innovation programme under the Marie Sk\l{}odowska-Curie grant agreement No. 839090. 
JB, DH, DAH, and CP were supported by NSF grant AST-1911225.
TMB was funded by the CONICYT PFCHA / DOCTORADOBECAS CHILE/2017-72180113.

This work has been partially supported by the Spanish grant PGC2018-095317-B-C21 within the European Funds for Regional Development (FEDER).

Based on observations made with the Nordic Optical Telescope, owned in collaboration by the University of Turku and Aarhus University, and operated jointly by Aarhus University, the University of Turku and the University of Oslo, representing Denmark, Finland and Norway, the University of Iceland and Stockholm University at the Observatorio del Roque de los Muchachos, La Palma, Spain, of the Instituto de Astrofisica de Canarias.

Observations from the NOT were obtained through the NUTS and NUTS2 collaboration which are supported in part by the Instrument Centre for Danish Astrophysics (IDA). The data presented here were obtained in part with ALFOSC, which is provided by the Instituto de Astrofisica de Andalucia (IAA) under a joint agreement with the University of Copenhagen and NOTSA. 

Based on observations made with the GTC telescope, in the Spanish Observatorio del Roque de los Muchachos of the Instituto de Astrofísica de Canarias, under Director’s Discretionary Time. 

This work has made use of data from the Asteroid Terrestrial-impact Last Alert System (ATLAS) project. ATLAS is primarily funded to search for near earth asteroids through NASA grants NN12AR55G, 80NSSC18K0284, and 80NSSC18K1575; by products of the NEO search include images and catalogues from the survey area. The ATLAS science products have been made possible through the contributions of the University of Hawaii Institute for Astronomy, the Queen’s University Belfast, the Space Telescope Science Institute, and the South African Astronomical Observatory.

The Liverpool Telescope is operated on the island of La Palma by Liverpool John Moores University in the Spanish Observatorio del Roque de los Muchachos of the Instituto de Astrofisica de Canarias with financial support from the UK Science and Technology Facilities Council.
This work makes use of data from the Las Cumbres Observatory network.\\

\noindent \textit{Facilities:} ATLAS; GTC; Las Cumbres Observatory; LT; NOT; Swift, WHT, ZTF.\\

\noindent \textit{Software:} \textsc{python} from \url{https://www.python.org/}, \textsc{photutils} \citep{Bradley19}, \textsc{astropy} \citep{Astropy18}, \textsc{iraf}, SNOoPY \citep{Cappellaro14}, FDSTfast, \textsc{floyds\_pipeline} \citep{Valenti14}, \textsc{george} \citep{Ambikasaran16}, \textsc{synow} \citep{Fisher00}, \textsc{mesa} \citep{Paxton11}, \textsc{snid} \citep{Blondin07}.  


\section*{Data Availability}

The data underlying this article are available in the appendix A (Tables A1 -- A5) and through the WISeREP (\url{https://wiserep.weizmann.ac.il/home}) archive \citep{Yaron12}.




\bibliographystyle{mnras}
\bibliography{Bibliography}

\section*{Affiliations}

$^{1}$ Finnish Centre for Astronomy with ESO (FINCA), FI-20014 University of Turku, Finland \\
$^{2}$ Tuorla Observatory, Department of Physics and Astronomy, FI-20014 University of Turku, Finland \\
$^{3}$ Department of Physics and Astronomy, University of Southampton, Southampton, SO17 1BJ, UK\\
$^{4}$ Facultad de Ciencias Astron\'omicas y Geof\'isicas, Universidad Nacional de La Plata, Paseo del Bosque S/N, B1900FWA, \\ La Plata, Argentina\\
$^{5}$ Instituto de Astrof\'isica de La Plata (IALP), CCT-CONICET-UNLP. Paseo del Bosque S/N, B1900FWA, La Plata, Argentina\\
$^{6}$ Kavli Institute for the Physics and Mathematics of the Universe (WPI), The University of Tokyo, 5-1-5 Kashiwanoha, \\  Kashiwa, Chiba 277-8583, Japan\\
$^{7}$ Universidad Nacional de Río Negro. Sede Andina, Mitre 630 (8400), Bariloche, Argentina.\\
$^{8}$ Consejo Nacional de Investigaciones Científicas y Tećnicas (CONICET), Argentina.\\
$^{9}$ INAF - Osservatorio Astronomico di Padova, Vicolo dell'Osservatorio 5, I-35122 Padova, Italy\\
$^{10}$ Departamento de Ciencias Fisicas, Universidad Andres Bello, Avda. Republica 252, Santiago, Chile\\
$^{11}$ Millennium Institute of Astrophysics (MAS), Nuncio Monse\~nor Sotero Sanz 100, Providencia, Santiago, Chile\\
$^{12}$ European Southern Observatory, Alonso de C\'ordova 3107, Casilla 19, Santiago, Chile\\
$^{13}$ Astrophysics Research Centre, School of Mathematics and Physics, Queens University Belfast, Belfast BT7 1NN, UK\\
$^{14}$ DTU Space, National Space Institute, Technical University of Denmark, Elektrovej 327, 2800 Kgs. Lyngby, Denmark \\
$^{15}$ School of Physics \& Astronomy, Cardiff University, Queens Buildings, The Parade, Cardiff, CF243AA, UK\\
$^{16}$ Institute of Space Sciences (ICE, CSIC), Campus UAB, Carrer de Can Magrans s/n, 08193 Barcelona, Spain\\
$^{17}$ School of Physics, O'Brien Centre for Science North, University College Dublin, Dublin, Ireland.\\
$^{18}$ Department of Physics and Astronomy, Aarhus University, Ny Munkegade, DK-8000 Aarhus C, Denmark\\ 
$^{19}$ Department of Physics, University of California, Santa Barbara, CA 93106-9530, USA\\
$^{20}$ Las Cumbres Observatory, 6740 Cortona Dr, Suite 102, Goleta, CA 93117-5575, USA\\
$^{21}$ Institute of Cosmology and Gravitation, University of Portsmouth, Portsmouth, PO1 3FX\\
$^{22}$ Departamento de F\'isica Te\'orica y del Cosmos, Universidad de Granada, E-18071 Granada, Spain\\
$^{23}$ Universit\'e de Lyon, F-69622, Lyon, France; Universit\'e de Lyon 1, Villeurbanne; CNRS/IN2P3, Institut de Physique des Deux Infinis, Lyon\\

\appendix

\section{Tables}
\label{ap1}

Photometry and spectroscopic observations of SN~2019cad.

\renewcommand{\thetable}{A\arabic{table}}
\setcounter{table}{0}
\input{table_photoATLAS.tex} 

\renewcommand{\thetable}{A\arabic{table}}
\setcounter{table}{1}
\input{table_photo.tex} 

\renewcommand{\thetable}{A\arabic{table}}
\setcounter{table}{2}
\input{table_photoNIR.tex} 

\renewcommand{\thetable}{A\arabic{table}}
\setcounter{table}{3}
\input{table_photoSwift.tex} 

\renewcommand{\thetable}{A\arabic{table}}
\setcounter{table}{4}
\input{table_photoZTF.tex} 

\renewcommand{\thetable}{A\arabic{table}}
\setcounter{table}{5}
\input{table_spectra.tex} 


\bsp	
\label{lastpage}
\end{document}

%% file: table_photoATLAS.tex
\begin{table}
\centering
\caption{ATLAS AB optical photometry.}
\label{photoatlas}
\begin{tabular}[t]{ccccccc}
\hline
\hline
UT date    &	MJD	& Phase           &  Band  & Magnitude  \\
           &            & (days)$^{\ast}$ &        &  (mag)     \\
\hline                                
\hline                                
20190227   &  58542.48  &    \nodata      &   $o$  & $>20.50$       \\
20190303   &  58546.46  &    \nodata      &   $o$  & $>20.50$       \\
20190307   &  58550.44  &    \nodata      &   $o$  & $>20.70$       \\
20190311   &  58554.42  &    3.88         &   $o$  & $19.91\pm0.11$ \\
20190315   &  58558.43  &    7.78         &   $o$  & $19.45\pm0.20$ \\
20190319   &  58562.39  &    11.64        &   $o$  & $18.43\pm0.13$ \\
20190323   &  58566.41  &    15.55        &   $o$  & $18.44\pm0.11$ \\
20190325   &  58568.41  &    17.50        &   $o$  & $18.63\pm0.06$ \\
20190329   &  58572.39  &    21.38        &   $o$  & $18.54\pm0.04$ \\
20190402   &  58576.43  &    25.31        &   $c$  & $19.45\pm0.05$ \\
20190404   &  58578.40  &    27.23        &   $o$  & $18.74\pm0.08$ \\
20190406   &  58580.40  &    29.18        &   $c$  & $19.43\pm0.10$ \\
20190408   &  58582.37  &    31.10        &   $o$  & $18.63\pm0.07$ \\
20190414   &  58588.39  &    36.96        &   $o$  & $17.64\pm0.09$ \\
20190416   &  58590.36  &    38.88        &   $o$  & $17.35\pm0.02$ \\
20190420   &  58594.34  &    42.76        &   $o$  & $17.33\pm0.03$ \\
20190422   &  58596.34  &    44.71        &   $o$  & $17.41\pm0.01$ \\
20190424   &  58598.32  &    46.63        &   $o$  & $17.46\pm0.02$ \\
20190426   &  58600.33  &    48.59        &   $o$  & $17.47\pm0.02$ \\
20190430   &  58604.34  &    52.50        &   $c$  & $18.04\pm0.03$ \\
20190506   &  58610.30  &    58.30        &   $o$  & $18.65\pm0.06$ \\
20190508   &  58612.27  &    60.22        &   $c$  & $18.76\pm0.05$ \\
20190510   &  58614.27  &    62.17        &   $o$  & $18.77\pm0.08$ \\
20190512   &  58616.27  &    64.12        &   $o$  & $18.71\pm0.08$ \\
20190514   &  58618.29  &    66.09        &   $o$  & $19.16\pm0.26$ \\
20190516   &  58620.31  &    68.05        &   $o$  & $19.34\pm0.15$ \\
20190518   &  58622.26  &    69.95        &   $o$  & $19.47\pm0.30$ \\
20190522   &  58626.24  &    73.83        &   $o$  & $19.40\pm0.13$ \\
20190524   &  58628.25  &    75.79        &   $o$  & $19.32\pm0.28$ \\
20190530   &  58634.25  &    81.63        &   $o$  & $19.57\pm0.23$ \\
20190601   &  58636.25  &    83.58        &   $c$  & $22.28\pm0.42$ \\
\hline
\end{tabular}
\begin{list}{}{}
\item \textbf{Notes:} 
\item $^{\ast}$ Rest-frame phase in days from explosion, MJD=$58550.44\pm4$.
\end{list}
\end{table}

%% file: table_photo.tex
\begin{table*}
\centering
\scriptsize
\caption{Optical photometry of SN~2019cad.}
\label{photo}
\begin{tabular}[t]{cccccccccccc}
\hline
\hline

UT date  &   MJD    & Phase           &     $u$        &      $B$	&      $V$	 &      $g$       &	$r$	   &	$i$	    &   $z$         & Telescope$^{\ddagger}$   \\

        &          & (days)$^{\ast}$ &     (mag)      &     (mag)      &     (mag)      &     (mag)      &     (mag)      &    (mag)       &   (mag)       &             \\
\hline
\hline                                              
20190322 & 58565.32 &	  14.49       & $20.10\pm0.10^{\star}$ & $20.05\pm0.09$ & $19.25\pm0.06$ & $19.43\pm0.04$ & $18.77\pm0.03$ & $18.73\pm0.03$ & \nodata & LCOGT \\
20190328 & 58571.11 &     20.13       &     \nodata    & $21.24\pm0.07$ & $19.27\pm0.03$ & $20.39\pm0.03$ & $18.84\pm0.03$ & $18.58\pm0.04$ &     \nodata   & LCOGT \\
20190329 & 58572.17 &     21.16       &     \nodata    & $21.49\pm0.07$ & $19.58\pm0.03$ & $20.51\pm0.05$ & $19.01\pm0.03$ & $18.77\pm0.03$ &     \nodata   & LCOGT \\
20190421 & 58594.85 &     43.26       & $18.44\pm0.02$ &  \nodata       &    \nodata    & $17.58\pm0.03$ & $17.32\pm0.02$ & $17.19\pm0.02$ & $17.05\pm0.02$ & LT    \\
20190423 & 58596.86 &     45.21       &   \nodata      &  \nodata       &    \nodata    &    \nodata     & $17.31\pm0.02$ &    \nodata     &    \nodata     & NOT   \\
20190424 & 58597.85 &     46.18       & $18.66\pm0.02$ &  \nodata       &    \nodata    & $17.63\pm0.02$ & $17.33\pm0.01$ & $17.14\pm0.01$ & $17.07\pm0.02$ & LT    \\
20190427 & 58600.85 &     49.10       & $18.69\pm0.02$ &  \nodata       &    \nodata    & $17.86\pm0.03$ & $17.46\pm0.01$ & $17.26\pm0.01$ & $17.16\pm0.02$ & LT    \\
20190428 & 58601.90 &     50.12       &   \nodata      &  \nodata       &    \nodata    &    \nodata     & $17.55\pm0.03$ &    \nodata     &    \nodata     & NOT   \\
20190429 & 58602.95 &     51.14       &   \nodata      &  \nodata       &    \nodata    & $18.22\pm0.01$ & $17.72\pm0.01$ & $17.56\pm0.01$ & $17.37\pm0.01$ & NOT   \\
20190430 & 58603.85 &     52.02       & $19.61\pm0.02$ &  \nodata       &    \nodata    & $18.16\pm0.02$ & $17.69\pm0.02$ & $17.48\pm0.02$ & $17.30\pm0.02$ & LT    \\
20190501 & 58604.92 &     53.06       &   \nodata      &  \nodata       &    \nodata    &    \nodata     & $17.98\pm0.08$ &    \nodata     &    \nodata     & NOT   \\
20190503 & 58606.85 &     54.94       &   \nodata      &  \nodata       &    \nodata    & $18.57\pm0.06$ & $17.93\pm0.03$ & $17.73\pm0.03$ & $17.55\pm0.03$ & LT    \\
20190511 & 58614.88 &     62.76       &   \nodata      &  \nodata       &    \nodata    &    \nodata     & $19.11\pm0.10$ &    \nodata     &    \nodata     & NOT   \\
20190513 & 58616.89 &     64.72       & $21.99\pm0.24$ &  \nodata       &    \nodata    & $20.03\pm0.03$ & $19.26\pm0.03$ & $18.84\pm0.03$ & $18.47\pm0.02$ & NOT   \\
20190516 & 58619.85 &     67.60       &   $>21.71$     &  \nodata       &    \nodata    & $20.19\pm0.09$ & $19.43\pm0.04$ & $19.05\pm0.03$ & $18.55\pm0.03$ & NOT   \\
20190603 & 58637.90 &     85.19       &   \nodata      & $22.23\pm0.10$ &    \nodata    & $21.50\pm0.06$ & $20.48\pm0.02$ & $20.06\pm0.04$ & $19.41\pm0.05$ & NOT   \\
20190606 & 58640.85 &     88.06       &   \nodata      &  \nodata       &    \nodata    &    \nodata     & $20.45\pm0.15$ &    \nodata     &    \nodata     & GTC   \\
20190611 & 58645.89 &     92.97       &   \nodata      &  \nodata       &    \nodata    & $22.01\pm0.07$ & $20.97\pm0.03$ & $20.51\pm0.06$ & $19.90\pm0.10$ & NOT   \\
20190619 & 58653.89 &     100.76      &   \nodata      &  \nodata       &    \nodata    & $>21.86$	 & $21.40\pm0.04$ & $20.92\pm0.10$ & $20.16 0.10$   & NOT   \\
20190622 & 58656.89 &     103.68      &   \nodata      &  \nodata       &    \nodata    & $>21.68$	 & $20.95\pm0.10$ &    \nodata     &    \nodata     & WHT   \\
\hline
\end{tabular}
\begin{list}{}{}
\item \textbf{Notes:} \\
$^{\ast}$ Rest-frame phase in days from explosion, MJD=$58550.44\pm4$.\\
$^{\ddagger}$ Telescope code: LCOGT: Las Cumbres Observatory; LT: 2.0-m Liverpool Telescope: NOT: Nordic Optical Telescope; GTC: Gran Telescopio Canarias; WHT: William Herschel Telescope. \\
$^{\star}$ Observation performed in the $U$-band. \\
$UBV$ photometry is in the Vega system, while $ugriz$ photometry is in the AB system.
\end{list}
\end{table*}

%% file: table_photoNIR.tex
\begin{table*}
\centering
\caption{$JHK$ Vega photometry of SN~2019cad obtained with NOTCam.}
\label{photonir}
\begin{tabular}[t]{cccccccccccc}
\hline
\hline
UT date  &   MJD    & Phase           &     $J$        &      $H$	&      $K$      \\ 
         &          & (days)$^{\ast}$ &     (mag)      &     (mag)      &     (mag)     \\ 
\hline
\hline                                              
20190430 & 58603.95 &    52.12        & $17.29\pm0.08$ & $16.20\pm0.10$ & $16.35\pm0.22$ \\
20190521 & 58624.89 &    72.51        & $18.00\pm0.11$ & $17.26\pm0.24$ & $17.74\pm0.32$ \\
20190613 & 58647.89 &    94.92        & $20.09\pm0.13$ & $18.40\pm0.15$ & $18.24\pm0.36$ \\
\hline
\end{tabular}
\begin{list}{}{}
\item \textbf{Notes:} \\
$^{\ast}$ Rest-frame phase in days from explosion, MJD=$58550.44\pm4$.
\end{list}
\end{table*}

%% file: table_photoSwift.tex
\begin{table}
\centering
\caption{UV photometry obtained with Swift in the AB system.}
\label{photoswift}
\begin{tabular}[t]{ccccccccc}
\hline
\hline
UT date    &	MJD	& Phase           &   UVW2    &    UVM2   &     UVW1  &      U    &      B    &     V     \\
           &            & (days)$^{\ast}$ &   (mag)   &    (mag)  &    (mag)  &    (mag)  &    (mag)  &   (mag)   \\
\hline                                
\hline                                
20190507   & 58611.14   &   59.12         & $>18.72$ & $>18.84$ & $>18.93$ & $>18.54$ & $>18.50$ & $>17.78$ \\
\hline
\end{tabular}
\begin{list}{}{}
\item \textbf{Notes:} 
\item $^{\ast}$ Rest-frame phase in days from explosion, MJD=$58550.44\pm4$.
\end{list}
\end{table}

%% file: table_photoZTF.tex
\begin{table}
\centering
\caption{ZTF AB optical photometry.}
\label{photoZTF}
\begin{tabular}[t]{ccccccc}
\hline
\hline
UT date    &	MJD	& Phase           &  Band  & Magnitude  \\
           &            & (days)$^{\ast}$ &        &  (mag)     \\
\hline                                
\hline   
20190314   &  58556.20  &     5.61       &   $r$  &  $>19.61$       \\
20190316   &  58558.20  &     7.56       &   $r$  &  $>19.18$       \\ 
20190317   &  58559.24  &     8.57       &   $r$  &  $19.02\pm0.11$ \\
20190317   &  58559.30  &     8.63       &   $g$  &  $19.31\pm0.24$ \\
20190320   &  58562.15  &     11.41      &   $r$  &  $18.75\pm0.14$ \\
20190320   &  58562.24  &     11.49      &   $g$  &  $>18.54$       \\
20190325   &  58567.20  &     16.32      &   $g$  &  $19.73\pm0.18$ \\
20190325   &  58567.22  &     16.34      &   $r$  &  $18.67\pm0.09$ \\
20190330   &  58572.20  &     21.19      &   $g$  &  $20.16\pm0.24$ \\
20190330   &  58572.22  &     21.21      &   $r$  &  $18.83\pm0.15$ \\
20190402   &  58575.21  &     24.13      &   $r$  &  $18.95\pm0.13$ \\
20190402   &  58575.26  &     24.17      &   $g$  &  $20.09\pm0.23$ \\
20190407   &  58580.21  &     29.00      &   $r$  &  $19.08\pm0.11$ \\
20190409   &  58582.20  &     30.93      &   $r$  &  $18.76\pm0.08$ \\
20190411   &  58584.22  &     32.90      &   $g$  &  $19.09\pm0.12$ \\
20190414   &  58587.21  &     35.81      &   $g$  &  $18.21\pm0.06$ \\
20190416   &  58589.22  &     37.77      &   $r$  &  $17.50\pm0.05$ \\
20190418   &  58591.34  &     39.84      &   $r$  &  $17.39\pm0.05$ \\
20190420   &  58593.33  &     41.77      &   $r$  &  $17.37\pm0.06$ \\
20190421   &  58594.16  &     42.58      &   $r$  &  $17.41\pm0.03$ \\
20190421   &  58594.18  &     42.60      &   $g$  &  $17.68\pm0.05$ \\
20190425   &  58598.20  &     46.52      &   $g$  &  $17.79\pm0.05$ \\
20190426   &  58599.15  &     47.44      &   $r$  &  $17.49\pm0.05$ \\
20190428   &  58601.26  &     49.50      &   $r$  &  $17.65\pm0.05$ \\
20190502   &  58605.22  &     53.36      &   $g$  &  $18.58\pm0.08$ \\
20190505   &  58608.19  &     56.25      &   $g$  &  $18.94\pm0.10$ \\
20190514   &  58617.19  &     65.01      &   $g$  &  $19.79\pm0.17$ \\
20190514   &  58617.22  &     65.04      &   $r$  &  $18.99\pm0.11$ \\
20190515   &  58618.26  &     66.06      &   $r$  &  $19.30\pm0.20$ \\
\hline
\end{tabular}
\begin{list}{}{}
\item \textbf{Notes:} 
\item $^{\ast}$ Rest-frame phase in days from explosion, MJD=$58550.44\pm4$.
\end{list}
\end{table}

%% file: table_spectra.tex
\begin{table*}
\centering
\caption{Spectroscopic observations of SN~2019cad}
\label{tspectra}
\begin{tabular}[t]{cccccccccc}
\hline
\hline
UT date   &	MJD   & Phase       	&  Range       &  Telescope   & Grism/Grating  \\
          &           & (days)$^{\ast}$ &  (\AA)       & +Instrument  &                \\
\hline
\hline
20190322  & 58564.36  &    13.56        & $3500-10000$ & LCOGT+FLOYDS &   red/blue     \\  
20190325  & 58567.95  &    17.51        & $3780-9200$  & P60+SEDM   &                \\ 
20190327  & 58569.32  &    18.39        & $3500-9250$  & LCOGT+FLOYDS &   red/blue     \\  
20190401  & 58573.38  &    22.34        & $3500-10000$ & LCOGT+FLOYDS &   red/blue     \\  
20190419  & 58592.02  &    40.50        & $4020-7990$  & LT+SPRAT     &   VPH          \\  
20190420  & 58593.95  &    42.38        & $4020-7990$  & LT+SPRAT     &   VPH          \\ 
20190423  & 58596.87  &    45.22        & $3600-9180$  & NOT+ALFOSC   &   Grism\#4     \\  
20190428  & 58601.91  &    50.13        & $3420-9690$  & NOT+ALFOSC   &   Grism\#4     \\  
20190501  & 58604.94  &    53.08        & $3500-9590$  & NOT+ALFOSC   &   Grism\#4     \\  
20190511  & 58614.90  &    62.78        & $3500-9690$  & NOT+ALFOSC   &   Grism\#4     \\  
20190516  & 58619.92  &    67.67        & $3500-9650$  & NOT+ALFOSC   &   Grism\#4     \\  
20190606  & 58640.93  &    88.14        & $4000-10235$ & GTC+OSIRIS   & R1000B+R1000R  \\  
\hline 
\end{tabular}
\begin{list}{}{}
\item \textbf{NOTES:}
\item $^{\ast}$ Rest-frame phase in days from explosion, MJD=$58550.44\pm4$.
\item \textbf{Telescope code:} LCOGT: Las Cumbres Observatory; P60: 60-inch Telescope; 
LT: 2.0-m Liverpool Telescope; NOT: Nordic Optical Telescope; GTC: Gran Telescopio Canarias
\end{list}
\end{table*}